\documentclass[journal]{IEEEtran}

\usepackage{url}
\usepackage{array}
\usepackage{cancel}
\usepackage{siunitx}
\usepackage{multirow}
\usepackage{graphicx}
\usepackage{textcomp}
\usepackage{multirow}
\usepackage{stfloats}
\usepackage{listings}
\usepackage{colortbl}
\usepackage{algorithm}
\usepackage[T1]{fontenc}
\usepackage[10pt]{moresize}
\usepackage[utf8]{inputenc}
\usepackage[noadjust]{cite}
\usepackage{mathtools, cuted}
\usepackage[noend]{algpseudocode}
\usepackage[table,x11names]{xcolor}
\usepackage{amsmath,bm,amssymb,amsfonts}

\definecolor{skyblue}{rgb}{0.53, 0.81, 0.92}
\definecolor{codegreen}{rgb}{0,0.6,0}
\definecolor{codegray}{rgb}{0.5,0.5,0.5}
\definecolor{codepurple}{rgb}{0.58,0,0.82}
\definecolor{backcolour}{rgb}{0.95,0.95,0.92}
\definecolor{LightCyan}{rgb}{0.88,1,1}
\definecolor{codeblue}{RGB}{49,49,255}
\definecolor{codeorange}{RGB}{255,143,102}
\definecolor{codewhite}{RGB}{255,255,255}
\definecolor{bittersweet}{rgb}{1.0, 0.44, 0.37}
\definecolor{columbiablue}{rgb}{0.61, 0.87, 1.0}
\definecolor{cornellred}{rgb}{0.7, 0.11, 0.11}

\newcommand {\ccsb} {\cellcolor{skyblue}}
\newcommand {\cccb} {\cellcolor{columbiablue}}

\lstdefinestyle{mystyle}{
	backgroundcolor=\color{backcolour}, 
	commentstyle=\color{codegreen},
	keywordstyle=\color{codepurple},
	numberstyle=\tiny\color{codegreen},
	stringstyle=\color{codepurple},
	identifierstyle=\color{cornellred},
	basicstyle=\ttfamily\footnotesize,
	breakatwhitespace=false,         
	breaklines=true,                 
	captionpos=t,                    
	keepspaces=true,                 
	numbersep=5pt,                  
	showspaces=false,                
	showstringspaces=false,
	showtabs=false,                  
	tabsize=2
}

\lstdefinestyle{plainstyle}{
	basicstyle=\ttfamily\footnotesize,
	keepspaces=true,                                
	numbersep=5pt,                  
	showspaces=false,                
	showstringspaces=false,
	showtabs=false,                  
	frame=tb,
	captionpos=b,
	tabsize=2
}

\IEEEoverridecommandlockouts

\graphicspath{{./figs/}}

\newlength{\textfloatsepsave} 
\begin{document}
	
	\title{Hybrid Protection of Digital FIR Filters}
	
	\author{Levent Aksoy,~\IEEEmembership{Member,~IEEE,} Quang-Linh~Nguyen,~\IEEEmembership{Student Member,~IEEE,} Felipe Almeida, Jaan Raik,~\IEEEmembership{Member,~IEEE,} Marie-Lise~Flottes,~\IEEEmembership{Member,~IEEE,} Sophie~Dupuis,~\IEEEmembership{Member,~IEEE,}
		and~Samuel~Pagliarini,~\IEEEmembership{Member,~IEEE}
		
		\thanks{This work has been partially conducted in the project ``ICT programme'' which was supported by the European Union through the European Social Fund. It was also partially supported by European Union's Horizon 2020 research and innovation programme under grant agreement No 952252 (SAFEST) and by the project MOOSIC ANR-18-CE39-0005 of the French National Research Agency (ANR).}
		
		\thanks{L.~Aksoy, F.~Almeida, J.~Raik, and S.~Pagliarini are with the Department of Computer Systems, Tallinn University of Technology, Tallinn, Estonia (e-mail: \{levent.aksoy, felipe.almeida, jaan.raik, samuel.pagliarini\}@taltech.ee.)}
		
		\thanks{Q.-L.~Nguyen, M.-L.~Flottes, and S.~Dupuis are with LIRMM, University of Montpellier, CNRS, Montpellier, France (e-mail: \{quang-linh.nguyen, marie-lise.flottes, sophie.dupuis\}@lirmm.fr.)}
	}
	
	\maketitle
	
	\begin{abstract}
		A digital Finite Impulse Response (FIR) filter is a ubiquitous block in digital signal processing applications and its behavior is determined by its coefficients. To protect filter coefficients from an adversary, efficient obfuscation techniques have been proposed, either by hiding them behind decoys or replacing them by key bits. In this article, we initially introduce a query attack that can discover the secret key of such obfuscated FIR filters, which could not be broken by existing prominent attacks. Then, we propose a first of its kind hybrid technique, including both hardware obfuscation and logic locking using a point function for the protection of parallel direct and transposed forms of digital FIR filters. Experimental results show that the hybrid protection technique can lead to FIR filters with higher security while maintaining the hardware complexity competitive or superior to those locked by prominent logic locking methods. It is also shown that the protected multiplier blocks and FIR filters are resilient to existing attacks. The results on different forms and realizations of FIR filters show that the parallel direct form FIR filter has a promising potential for a secure design.
	\end{abstract}
	
	
	\section{Introduction}

Due to the increase in the design complexity of Integrated Circuits (ICs) and the rising costs of chip fabrication at advanced technology nodes, the IC supply chain has become heavily specialized and globalized~\cite{dsb15}. Design houses have been combining their Intellectual Properties (IPs) with many others purchased from third-parties and resorting to \emph{untrusted} foundries for fabrication. Although such globalization reduces the overall cost of producing an IC, it leads to serious security threats -- especially for IPs -- such as piracy, overuse, modification, and reverse engineering~\cite{amir17}. Over the years, IP protection has received a significant amount of interest and efficient methods, including watermarking~\cite{kahng98}, digital rights management~\cite{alkabani07}, metering~\cite{koushanfar07}, and hardware obfuscation~\cite{hoque20}, have been introduced. Among these techniques, only hardware obfuscation can prevent IP theft, while the others are useful to prove the IP owner and reveal the IP owner's rights during a litigation process. Hardware obfuscation aims to make the design less clear and hard to understand for an adversary, by hiding the design content using structural transformations, locking the design functionality using additional logic with key bits, and exploiting camouflaged gates~\cite{hoque20}.

Digital filtering is frequently used in Digital Signal Processing (DSP) applications and Finite Impulse Response (FIR) filters are generally preferred due to their stability and linear phase property~\cite{wanhammar99}. Since filter coefficients determine the filter behavior, they are actually an IP and need protection from reverse engineering by an adversary. Although there exist many efficient high-level and behavioral obfuscation methods proposed for protecting IPs~\cite{muttaki22, pilato21, islam20, pilato18, sengupta17, chakraborty10}, digital FIR filters require specialized obfuscation techniques, since they should behave according to their specifications, such as passband and stopband frequencies and ripples~\cite{yu10}. However, there exist only a limited number of techniques proposed to obfuscate DSP circuits and especially, digital filters~\cite{lao15, bottegal, aksoy21, aksoy22}.  The technique of~\cite{lao15} generates the desired filter and also its obfuscated versions, grouped in two categories as meaningful and unmeaningful in terms of filter behavior, using high-level transformations, and combines these realizations using a key-based finite state machine and a reconfigurator. To make the reverse engineering of coefficients harder for an \mbox{end-user}, adding input and output noises was proposed in~\cite{bottegal}. Recently, we introduced a hardware obfuscation technique that hides the filter coefficients behind decoys~\cite{aksoy21,aksoy22}. In~\cite{aksoy21}, decoys can be selected based on their Hamming distance to reduce the hardware complexity or chosen randomly to increase the corruption at the filter output. Since an obfuscated FIR filter may still generate the desired behavior under a wrong key in~\cite{aksoy21}, decoys are selected in such a way that the obfuscated filter presents the desired behavior only when the secret key is provided in~\cite{aksoy22}. To do so, the lower and upper bounds of each filter coefficient are found and decoys are selected beyond these bounds. In~\cite{aksoy21,aksoy22}, the folded design of an FIR filter is considered as a case study and its Time-Multiplexed Constant Multiplication (TMCM) block is obfuscated at Register-Transfer Level (RTL).

In this article, we initially introduce the query attack, which can discover the original filter coefficients hidden behind decoys~\cite{aksoy21,aksoy22} or replaced by key bits~\cite{pilato21}. Then, we propose a hybrid technique, which includes both hardware obfuscation and logic locking, for the protection of digital FIR filters. To do so, first, we describe a defense technique that obfuscates the multiplier blocks of parallel direct and transposed forms of an FIR filter, i.e., Constant Array Vector Multiplication (CAVM) and Multiple Constant Multiplication (MCM), respectively, using decoys. We also present their \mbox{hardware-efficient} realizations with and without multipliers. Second, we enhance this obfuscation technique by locking the obfuscated design using a point function to make the protected design resilient to well-known attacks and by thwarting the query attack to determine the secret key. The hybrid protection technique works at RTL and can be easily adapted to any application including constant multiplications, such as image and video processing and neural networks. The main contributions of this article are as follows:
\begin{itemize}
	\item {Query attack developed for breaking designs generated by constant obfuscation techniques;}
	\item {Secure hybrid technique, consisting of hardware obfuscation and logic locking, developed for the protection of FIR filters with different forms and realizations;}
	\item {Comprehensive results on obfuscation and logic locking of FIR filters in terms of hardware complexity, attack resiliency, and filter behavior.}
\end{itemize}

Experimental results clearly show that the proposed hybrid protection technique leads to FIR filter designs with higher security and competitive hardware complexity when compared to previously proposed hardware obfuscation and logic locking methods. As an interesting outcome of this work, we show that the parallel direct form filter has better resiliency properties than other FIR filter forms and realizations. 

The remainder of this article is organized as follows: Section~\ref{sec:background} presents background concepts. The query attack is described in Section~\ref{sec:attack} and the hybrid protection method is introduced in Section~\ref{sec:defense}. Experimental results are presented in Section~\ref{sec:results}. Further discussions on how other techniques may identify the original filter coefficients are given in Section~\ref{sec:discussion}. Finally, Section~\ref{sec:conclusions} concludes the article.

	\section{Background}
\label{sec:background}

This section initially presents frequently used notations and then, gives details on digital FIR filters and multiplierless constant multiplications. Finally, it summarizes related work.

\subsection{Notations}

Table~\ref{tab:notations} presents notations of important parameters used in the description of obfuscation and logic locking techniques.

\begin{table}[h]
	\centering
	\scriptsize
	\vspace*{-2mm}
	\caption{Summary of Notations}
	\vspace*{-2mm}
	\begin{tabular}{c|l}
		$c$ & Constant/filter coefficient\\
		$C$ & Constant array/set\\
		$n$ & Number of constants/filter coefficients \\
		$mbw$ & Maximum bit-width of constants/filter coefficients\\
		$X$ & Input variable/filter input \\
		$ibw$ & Bit-width of the input variable/filter input\\
		$Y$ & Output variable/filter output \\
		$k$ & Key bit \\
		$\bm{K}$ & Secret key \\
		$p$ & Total number of key bits \\
		$v$ & Number of key bits for obfuscation \\
		$w$ & Number of key bits for logic locking \\
	\end{tabular}
	\vspace*{-2mm}
	\label{tab:notations}
\end{table}

\subsection{Digital FIR Filters}

\begin{figure}[t]
	\centerline{\includegraphics[width=8.0cm]{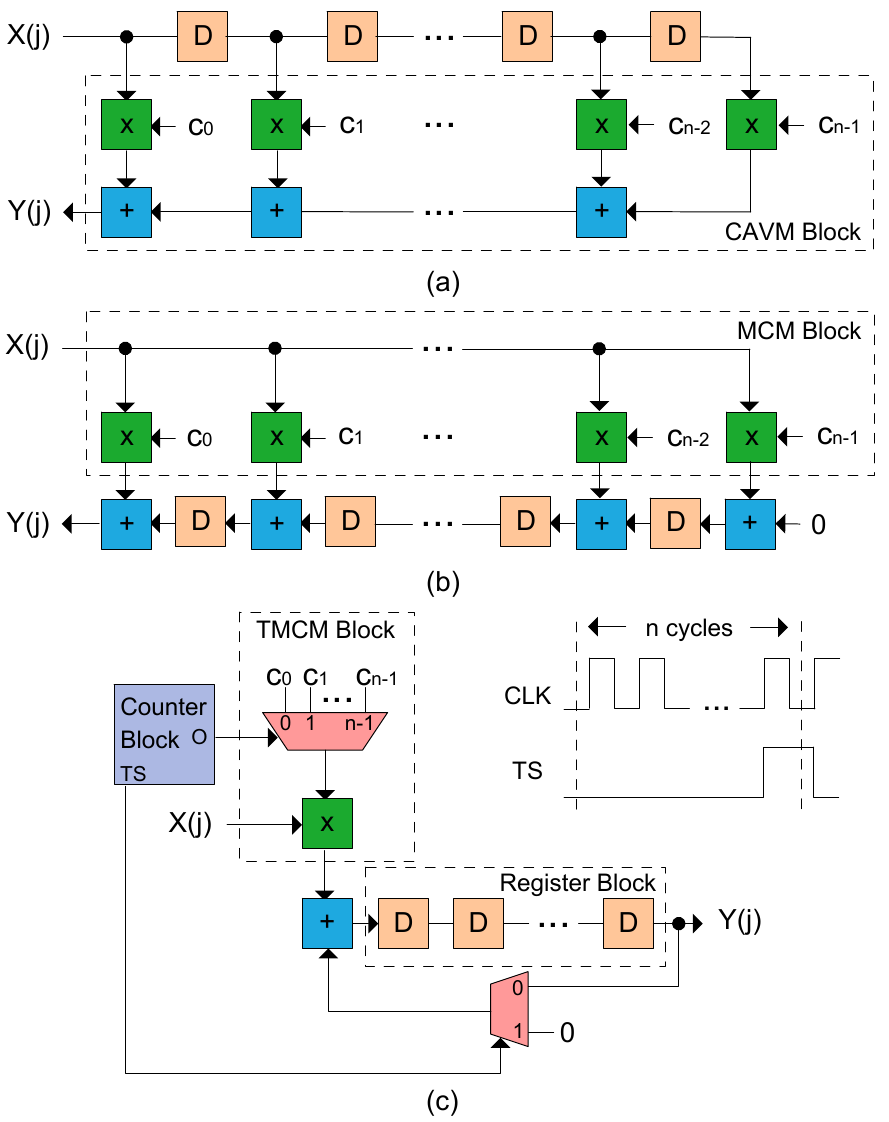}}
	\vspace*{-2mm}
	\caption{Designs of an FIR filter: (a)~parallel direct form; (b)~parallel transposed form; (c)~folded transposed form, where the counter counts from 0 to $n-1$.}
	\label{fig:directfir}
	\vspace*{-6mm}
\end{figure}

The FIR filter output $Y(j)$ is given as $\sum_{i=0}^{n-1}c_i\cdot X(j-i)$, where $n$ is the filter length, $c_i$ is the $i^{th}$ filter coefficient, and $X(j-i)$ is the $i^{th}$ previous filter input with $0 \leq i \leq n-1$. Fig.~\ref{fig:directfir} shows the parallel and folded realizations of an FIR filter. Note that the filter output is obtained in a single clock cycle in a parallel design, as shown in the direct and transposed forms in Figs.~\ref{fig:directfir}(a)-(b). On the other hand, the folded realization leads to a design with the least hardware complexity, since the common operations are re-used. However, it requires $n$ clock cycles to compute the filter output, as shown in Fig.~\ref{fig:directfir}(c). 

\begin{figure*}[t]
	\centerline{\includegraphics[width=16.0cm]{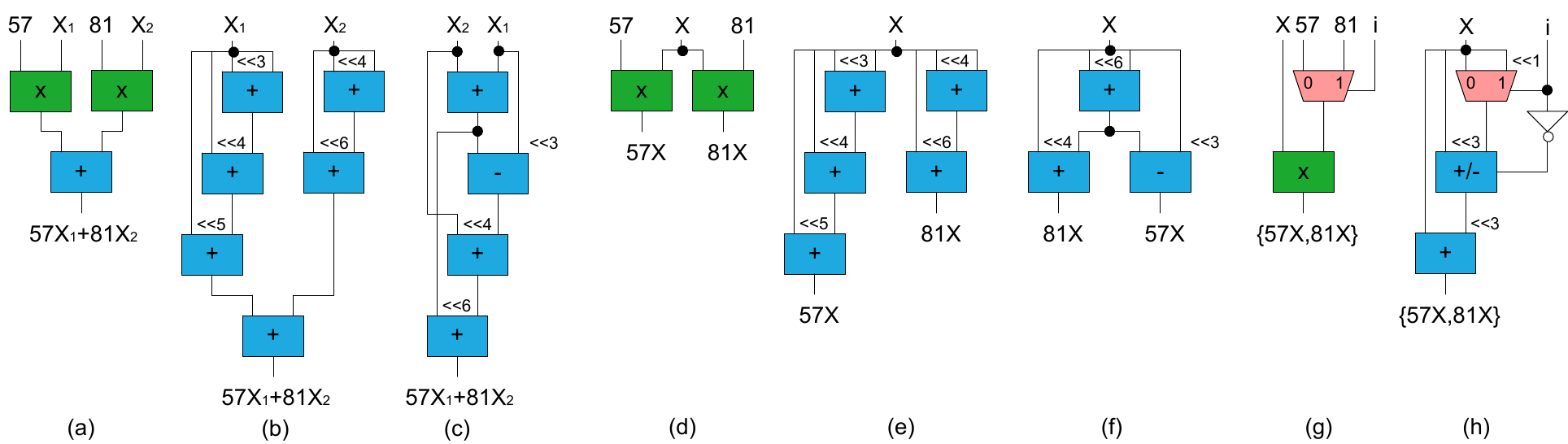}}
	\vspace*{-2mm}
	\caption{Realizations of the CAVM (a-c), MCM (d-f), and TMCM (g-h) blocks including constants 57 and 81: (a)~using multipliers; (b)~the DBR method~\cite{ercegovac03}; (c)~the method of~\cite{aksoy14_cavm}; (d)~using multipliers; (e)~the DBR method~\cite{ercegovac03}; (f)~the method of~\cite{aksoy10}; (g)~using a multiplier; (h)~the method of~\cite{aksoy14_tmcm}.}
	\label{fig:cavm}
	\vspace*{-2mm}
\end{figure*}

\begin{figure*}[t]
	\centerline{\includegraphics[width=17.0cm]{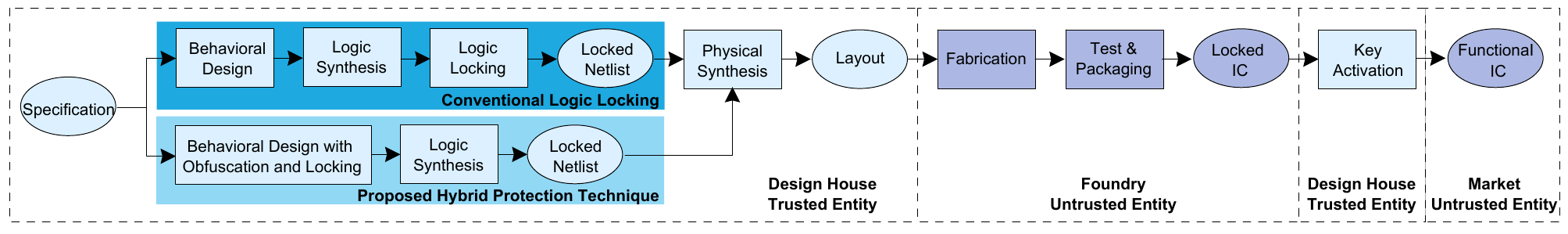}}
	\vspace*{-2mm}
	\caption{Conventional logic locking and proposed hybrid protection technique in the IC design flow (adapted from~\cite{yasin17}).}
	\label{fig:icflow}
	\vspace*{-6mm}
\end{figure*}

\subsection{Multiplierless Design of Constant Multiplications}

Multiplication of constant(s) by variable(s) is a ubiquitous and crucial operation in many DSP applications. Among others presented in~\cite{aksoy14_tutorial}, the CAVM, MCM, and TMCM blocks can be used in the design of a filter, as shown in Fig.~\ref{fig:directfir}. They are defined as follows:

\begin{enumerate}
	\item {The \textit{CAVM operation} implements the multiplication of a \(1 \times n\) constant array \(C\) by an \(n \times 1\) input vector \(X\), i.e., \(Y = \sum_{i}c_iX_i\) with \(1 \leq i \leq n\).}
	\item {The \textit{MCM operation} computes the multiplication of a set of \(n\) constants \(C\) by a single variable \(X\), i.e., \(Y_i=c_iX\) with \(1 \leq i \leq n\).}
	\item {The \textit{TMCM operation} realizes the multiplication of a constant selected from a set of \(n\) constants \(C\) by a single variable \(X\) at a time, i.e., \(Y = c_{i}X\) with \(1 \leq i \leq n\).}
\end{enumerate}

Since the constants are determined beforehand, these constant multiplications can be realized using addition, subtraction, and shift operations under the shift-adds architecture. Note that parallel shifts can be implemented virtually for free in hardware using only wires. A straightforward shift-adds design technique, called the Digit-Based Recoding (DBR)~\cite{ercegovac}, can realize constant multiplications in two steps: i)~define the constants under a particular number representation, e.g., binary; ii)~for the nonzero digits in the representation of constants, shift the input variables according to digit positions and add/subtract the shifted variables with respect to digit values. Furthermore, the number of operations can be reduced by maximizing the sharing of common subexpressions among constant multiplications~\cite{voronenko07, aksoy10, tummeltshammer07, aksoy14_tmcm, aksoy14_cavm}. 

As a simple example, consider the CAVM, MCM, and TMCM blocks realizing constant multiplications, where $C$ includes $57 = (111001)_{bin}$ and $81 = (1010001)_{bin}$. These constant multiplications are shown in Fig.~\ref{fig:cavm}. Note that the adder/subtractor shown in Fig.~\ref{fig:cavm}(h) behaves as an adder or a subtractor when its select input is 0 or 1, respectively. Observe from Figs.~\ref{fig:cavm}(b)-(c) and~(e)-(f) that the sharing of common subexpressions can lead to a significant reduction under the shift-adds architecture in terms of the number of operations with respect to the DBR method. 

\subsection{Related Work}

Hardware obfuscation can take place at different stages in the IC design flow, e.g., high-level synthesis~\cite{pilato18}, RTL~\cite{pilato21}, gate-level~\cite{roy08}, and layout level~\cite{raj13}. In hardware obfuscation, locking the design functionality is a common practice. Fig.~\ref{fig:icflow} presents conventional logic locking applied at gate-level in the IC design flow. Note that after the layout of the locked netlist is shipped to the foundry without revealing the secret key, the locked IC is produced and delivered back to the design house. Then, values of the secret key are stored in a \mbox{tamper-proof} memory and the functional IC is sent to the market.

\subsubsection{Defenses}

Earlier logic locking methods have been applied at \mbox{gate-level}. After the introduction of the concept of Random Logic Locking (RLL) using {\sc xor/xnor} gates in~\cite{roy08}, many works focused on different types of key logic, such as {\sc and/or}, multiplexors (MUXes), and look-up tables, taking into account the hardware complexity of the locked circuit~\cite{dupuis19}. However, the satisfiability (SAT)-based attack~\cite{subramanyan15} overcame all the defenses existing at that time. To thwart the \mbox{SAT-based} attack and its variants, circuits have been locked using a point function that forces these attacks to explore an exponential number of queries~\cite{yasin16, yasin17, xie19, shakya19, sengupta20, nguyen21}. Moreover, the obfuscation of a locked design is considered in~\cite{zhou19}.

However, as mentioned in~\cite{muttaki22}, at a higher level in the IC design flow, the selection of critical blocks of the design to be obfuscated gets easier, the exploration of tradeoffs between overhead and attack resiliency becomes more efficient, and the optimization of the obfuscated design is more effective. Recently, high-level and behavioral obfuscation techniques have been presented in~\cite{sengupta17, pilato18, islam20, pilato21, muttaki22}. Related to digital FIR filters including a large number of constants, filter coefficients are obfuscated by replacing their bits by key bits in~\cite{pilato18, pilato21}. 

We note that our proposed hybrid protection technique works at one level higher than the gate-level, i.e., at RTL, as also shown in Fig.~\ref{fig:icflow}. 

\subsubsection{Attacks}

In logic locking, there are generally two threat models, namely oracle-less (OL) and oracle-guided (OG). In the OL threat model, only the gate-level netlist of the locked circuit is available to an adversary. In the OG threat model, it is assumed that an adversary can also obtain the functional IC programmed with the secret key from the market and use it as an oracle to apply inputs and observe outputs. Hence, in this model, the adversary has both the netlist of the locked circuit and the functional IC.

Under the OL threat model, due to the limited information available to the adversary, patterns in the structure of the locked netlist are studied using statistical analysis, Automated Test Pattern Generation (ATPG), and machine learning~\cite{chakraborty18, li19, alaql21, alrahis21}. Structural attacks, which identify and remove the logic inserted by a logic locking method, are proposed in~\cite{sirone19, yang19, zhaokun21}. 

\begin{algorithm}[t]
	\small
	\caption{The SAT-based attack~\cite{subramanyan15}}
	\begin{algorithmic}[1]
		\Statex \textbf{Inputs:} Locked circuit \textit{LC} and \textit{oracle}.
		\Statex \textbf{Output:} Secret key $\bm{K}$.
		\State $i := 1$ \Comment{Number of iterations}
		\State $F_1 = LC(X, K_1, Y_1) \land LC(X, K_2, Y_2)$
		\While {$sat[F_i \land (Y_1 \neq Y_2)]$}
		\State $X_i^{d} := sat\_assignment_X[F_i \land (Y_1 \neq Y_2)]$
		\State $Y_i^{d} : = oracle(X_i^{d})$
		\State $F_{i+1} := F_i \land LC(X_i^{d}, K_1, Y_i^{d}) \land LC(X_i^{d}, K_2, Y_i^{d}) $
		\State $i := i + 1$
		\EndWhile
		\State $\bm{K} := sat\_assignment_{K_1}(F_i)$
	\end{algorithmic}
	\label{algo:satattack}
\end{algorithm}

Under the OG threat model, the ATPG-based attack of~\cite{raj12} leverages testing principles, such as justification and sensitization while finding the secret key. The SAT-based attack~\cite{subramanyan15} iteratively finds Distinguishing Input Patterns (DIPs) that rule out wrong keys and achieves decryption as shown in Algorithm~\ref{algo:satattack}. It generates two locked circuits with the same inputs ($X$), but two different keys ($K_1$ and $K_2$) described in a Conjunctive Normal Form (CNF) formula in a SAT problem (line 2). Then, it finds a DIP, which generates different outputs on these circuits, using a SAT solver (line 4) and computes the output based on the found DIP using the oracle (line 5). It adds the Boolean equations including key bits into the SAT problem, which are obtained after inserting the values of these inputs and outputs into these circuits (line 6). This process is iterated until the SAT problem becomes unsatisfiable (line 3), meaning that there exists no DIP to distinguish wrong keys from the secret key. Finally, it determines the secret key as the one found in the last iteration (line 8). 

In a similar fashion, the \mbox{SAT-based} attack of~\cite{shen17} eliminates at least 2 DIPs in a single iteration. A Satisfiability Modulo Theory (SMT) solver is used instead of a SAT solver, providing more flexibility while encoding the problem~\cite{azar18, karfa20}. The so-called approximate attack of~\cite{shamsi17} aims for approximate functional recovery. The \mbox{SAT-based} attack of~\cite{shamsi19} achieves sequential deobfuscation using dynamic simplifications of key conditions. The attack of~\cite{xu17} discovers the vulnerabilities of the SAT-resilient logic locking methods of~\cite{yasin16, xie19}. In~\cite{limaye21}, a generic framework is developed to attack compound locking techniques. A security diagnosis tool, which can evaluate the structural vulnerability of a design locked by a provably secure logic locking technique, is introduced in~\cite{limaye22}. 

	\section{The Query Attack}
\label{sec:attack}

The SAT-based attack~\cite{subramanyan15} presented in Algorithm~\ref{algo:satattack} guarantees that the found values of \textbf{all} key bits are equal to those of the secret key. To do so, it may use a large number of queries that are required to eliminate all the wrong keys. On the other hand, our query attack proves that the found value of a \textbf{single} key bit is equal to that of the associated one in the secret key. To do so, it uses a small number of queries that make each key bit observable at a primary output. Hence, it slightly increases the SAT problem size when compared to the \mbox{SAT-based} attack. Thus, it can easily cope with circuits including a large number of gates and key bits~\cite{tan20} and logic structures, such as a multiplier and a tree of {\sc and} gates~\cite{subramanyan15}, which the SAT-based attack generally finds hard to handle. In this section, we initially describe the proposed query attack and then, present its results on obfuscated designs.

\subsection{Description}

Our proposed OG SAT-based query attack is described in Algorithm~\ref{algo:queryattack}. It initially finds queries using two strategies (line 1). In the first one, an ATPG tool is used to find the test patterns for the stuck-at-fault of each key bit on the locked circuit and the values of the related primary inputs are stored as queries. The aim of this strategy is to find input patterns that can propagate each key bit to a primary output, making it observable. In the second one, queries are obtained randomly. The aim of this strategy is to find input patterns that may make multiple key bits observable at primary outputs. In our experiments, we generate a total of $2p$ queries, where $p$ denotes the total number of key bits. 

\begin{algorithm}[t]
	\small
	\caption{The query attack}
	\begin{algorithmic}[1]
		\Statex \textbf{Inputs:} Locked circuit \textit{LC} and \textit{oracle}.
		\Statex \textbf{Output:} Proven values of the secret key $\bm{K}$.
		\State $Q := find\_queries(LC)$
		\State $F = LC(X, K, Y)$
		\For{$i$ := 1 to $2p$}
		\State $Y_i := oracle(Q_i)$
		\State $F := F \land LC(Q_i, K, Y_i)$
		\EndFor
		\State $K := sat\_assignment_{K}(F)$
		\For {$i$ := 0 to $p-1$} 
		\If {$unsat[F \land \overline{K_i}]$}
		\State $\bm{K_i} = K_i$
		\EndIf
		\EndFor
		\For{$i$ := 0 to $p-2$}
		\For{$j$ := $i+1$ to $p-1$}
		\If{$undefined(\bm{K_i})~\&~undefined(\bm{K_j})$}
		\If{$unsat[F \land (K_i \neq K_j)]$}
		\State $\bm{K_i} = \bm{K_j}$
		\ElsIf {$unsat[F \land (K_i \neq \overline{K_j})]$}
		\State $\bm{K_i} = \overline{\bm{K_j}}$
		\EndIf
		\EndIf
		\EndFor
		\EndFor
	\end{algorithmic}
	\label{algo:queryattack}
\end{algorithm}

Then, the locked circuit is described in a CNF formula $\mathbb{F}$ by expressing each gate in its CNF (line 2). For each query (lines 3-5), it is applied to the oracle and the values of primary outputs are obtained (line 4). Then, the related input and output values are assigned to the associated nets in the locked circuit, the constant values of these nets are propagated, and the Boolean equations including key bits are derived in a CNF formula and added into $\mathbb{F}$ (line 5). 

After all the queries are considered, the SAT problem $\mathbb{F}$ is solved using a SAT solver and the values of key bits are determined (line 6). Note that the locked circuit with the found values of key bits behaves exactly the same as the oracle under the given queries, but not under all possible input values. Hence, the found key is not guaranteed to be the secret key. 

However, the found value of a key bit can be proven correct by using the concept of \textit{proof by contradiction}. To do so, for each key bit (lines 7-9), the complement of its found value is added into $\mathbb{F}$ and the SAT solver is run. If there exists no solution to $\mathbb{F}$, i.e., the SAT problem is unsatisfiable, the value of the related key bit in the secret key is proven to be the one in the found solution. 

\begin{figure*}[t]
	\centering
	\includegraphics[width=16.0cm]{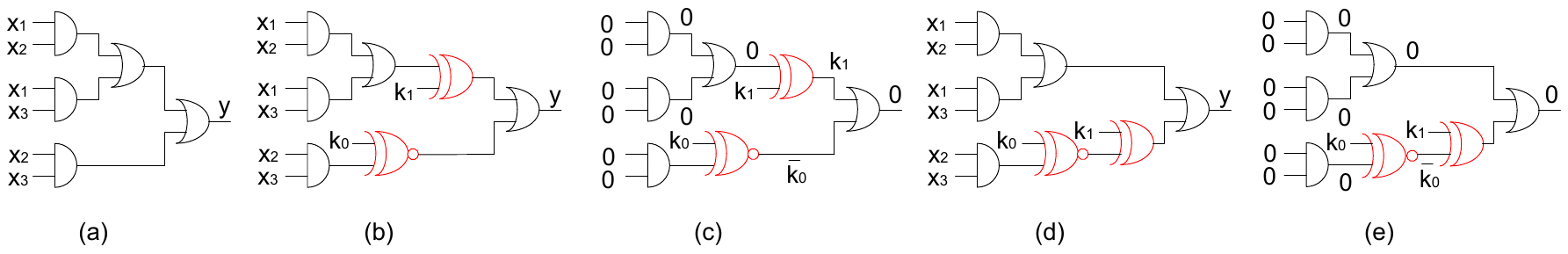}
	\vspace*{-2mm}
	\caption{Examples on the query attack: (a)~majority circuit; (b)-(c)~a locked majority circuit; (d)-(e)~another locked majority circuit.}
	\label{fig:majority}
	\vspace*{-6mm}
\end{figure*}

As an example, consider the majority circuit in Fig.~\ref{fig:majority}(a) and suppose that it is locked using {\sc xor/xnor} gates as given in Fig.~\ref{fig:majority}(b). Assume that a query is found as $x_1x_2x_3=000$ and thus, the value of its output $y$ is obtained as 0 using the oracle. After propagating these values on the locked circuit, a Boolean equation $\overline{k_0} \lor k_1 = 0$, i.e., $k_0 \land \overline{k_1}$ in CNF, is obtained as shown in Fig.~\ref{fig:majority}(c). In the SAT solution, the key bit values are found as $k_1k_0 = 01$. Note also that these are the proven key values since a SAT solver guarantees that there exists no solution to the SAT problem $\mathbb{F}$ when it is extended by either the constraint $k_0=0$, i.e., $\overline{k_0}$ in CNF or $k_1=1$, i.e., $k_1$ in CNF, due to a conflict with the found Boolean equation, i.e., $k_0 \land \overline{k_1}$ in CNF. 

We note that the query attack is also capable of proving if the value of a key bit, $k_i$, is equal to the value of another key bit, $k_j$, or its opposite (lines 10-16). To do so, we extend the SAT problem with $k_i \neq k_j$, i.e., $(k_i \lor k_j) \land (\overline{k_i} \lor \overline{k_j})$ in CNF, and $k_i \neq \overline{k_j}$, i.e., $(k_i \lor \overline{k_j}) \land (\overline{k_i} \lor k_j)$ in CNF, respectively, where $i \neq j$ and $0 \leq i,j \leq p-1$. We run the SAT solver and check if the SAT problem is unsatifiable. In this case, relations between two key bits are found independent of their values.

Returning back to our majority circuit, consider its another locked version given in Fig.~\ref{fig:majority}(d). Assume that a query is again found as $x_1x_2x_3 = 000$ and hence, the output $y$ is computed as 0. Thus, after the propagation of input and output values as shown in Fig.~\ref{fig:majority}(e), a Boolean equation $\overline{k_0} \oplus k_1 = 0$, i.e., $(k_0 \lor k_1) \land (\overline{k_0} \lor \overline{k_1})$ in CNF, is found. In the SAT solution, the key bit values are found as $k_1k_0 = 10$. Although the actual values of key bits could not be proven, it is found that $k_0$ and $k_1$ have opposite values after the SAT problem is extended with the Boolean equation $k_0 \neq \overline{k_1}$ and it becomes unsatisfiable. Hence, the values of key bits $k_1k_0 = 10$ or $k_1k_0 = 01$ in the locked design lead to the original majority circuit.

\begin{table}[t]
	\centering
	\caption{Details on FIR Filters and their TMCM Blocks.} 
	\vspace*{-2mm}
	\begin{tabular}{|l|cc|cc|}
		\hline
		\multicolumn{1}{|c|}{\multirow{2}{*}{Filter}} & \multicolumn{2}{c|}{Filter Details} & \multicolumn{2}{c|}{TMCM Details} \\
		\cline{2-5}
		& $n$ & $mbw$ & $\#in$ & $\#out$ \\
		\hline
		Mirzaei10a & 71  & 15 & 39 & 47 \\
		LimYu07    & 121 & 14 & 39 & 46 \\
		Mirzaei10b & 151 & 15 & 40 & 47 \\
		\hline
	\end{tabular}
	\label{tab:query_firs}
	\vspace*{-4mm}
\end{table}

\begin{table}[t]
	\centering
	\caption{Attacks on obfuscated tmcm blocks.} 
	\vspace*{-2mm}
	\begin{tabular}{|@{\hskip3pt}l@{\hskip3pt}|c|l|cccc|}
		\hline
		\multicolumn{1}{|c|}{\multirow{3}{*}{Filter}} & \multirow{3}{*}{$p$} & \multicolumn{1}{c|}{\multirow{3}{*}{Architecture}} & \multicolumn{4}{c|}{Attacks} \\
		\cline{4-7}
		& & & SAT & ATPG & \multicolumn{2}{c|}{Query} \\
		\cline{4-7}
		& & & time & time & prv & time \\
		\hline
		\multirow{2}{*}{Mirzaei10a} & \multirow{2}{*}{128} & {\sc tmcm-mul}~\cite{aksoy21}    & OoT & OoT & 128 & 212 \\
		&                      & {\sc tmcm-crk}~\cite{pilato21}   & OoT & OoT & 128 & 341 \\
		\hline
		\multirow{2}{*}{LimYu07}    & \multirow{2}{*}{128} & {\sc tmcm-mul}~\cite{aksoy21}    & OoT & OoT & 128 & 112 \\
		&                      & {\sc tmcm-crk}~\cite{pilato21}   & OoT & OoT & 128 & 365 \\
		\hline
		\multirow{2}{*}{Mirzaei10b} & \multirow{2}{*}{256} & {\sc tmcm-mul}~\cite{aksoy21}    & OoT & OoT & 256 & 809 \\
		&                      & {\sc tmcm-crk}~\cite{pilato21}   & OoT & OoT & 256 & 852 \\
		\hline
	\end{tabular}
	\label{tab:query}
	\vspace*{-6mm}
\end{table}

\subsection{Results}
\label{subsec:query_results}

First, three FIR filters with a large number of coefficients and a large bit-width of filter input and coefficients were used to demonstrate the performance of the query attack. They were taken from~\cite{firsuite}. Table~\ref{tab:query_firs} presents their details, where $n$ and $mbw$ are the number and maximum bit-width of coefficients, respectively. The folded realization of these filters was considered. Table~\ref{tab:query_firs} presents details on their TMCM blocks, where \textit{\#in} and \textit{\#out} are respectively their number of inputs and outputs when the \mbox{bit-width} of the input variable, i.e., $ibw$, is set to  32. These TMCM blocks were obfuscated using decoys under the architecture including MUXes and a multiplier~\cite{aksoy21}, denoted as \mbox{\sc tmcm-mul}, and also obfuscated by replacing constants with key bits~\cite{pilato21}, denoted as \mbox{{\sc tmcm-crk}}.

\begin{figure}[t]
	\centering
	\vspace*{-4mm}
	\parbox{8.5cm}{\vspace*{-2mm} \includegraphics[width=8.5cm]{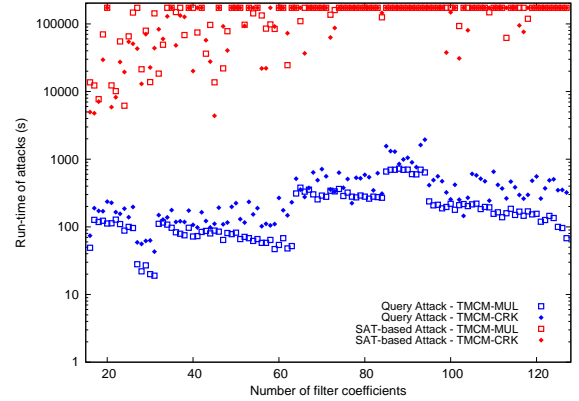}}
	\vspace*{-6mm}
	\caption{Run-time of attacks on obfuscated TMCM blocks.}  
	\vspace*{-6mm}
	\label{fig:attacks_random}
\end{figure}

Table~\ref{tab:query} presents the number of key bits $p$ and the results of the query attack along with the SAT- and ATPG-based attacks taken from~\cite{host15}. In this table, \textit{time} denotes the run-time of an attack in seconds and \textit{prv} stands for the number of key bits, whose values are proven by the query attack. Also, \textit{OoT} indicates that an attack could not find a solution due to the time limit, which was set to 2 days. The attacks were run on a computing cluster including Intel Xeon processing units at 2.4~GHz with 40 cores and 96~GB memory. The query attack was developed in Perl and equipped with the ATPG tool Atalanta~\cite{atalanta} and the SAT solver CaDiCaL~\cite{biere20}. It is available at \textit{https://github.com/Centre-for-Hardware-Security/}.

Observe from Table~\ref{tab:query} that the query attack can easily find the secret key of obfuscated designs while it is hard for the well-known attacks to find a solution. The main reason is that the TMCM block includes a multiplier block, where one of its inputs is the 32-bit input variable, and the \mbox{SAT-based} attack is not effective on designs including a multiplier as mentioned in~\cite{subramanyan15}. However, the query attack can deal with a small number of queries, which are sufficient to determine the value of each key bit, using a little computational effort.

Second, we generated a total of 112 FIR filters, where $n$ ranges between 16 and 127 when \textit{mbw} was set to 12, to find the impact of the number of constants and key bits on the performance of the query attack. Again, the folded design of these filters were considered and \textit{ibw} was set to 32. The TMCM blocks were obfuscated using $2^{\lfloor log_2n \rfloor + 1}$ key bits under the \mbox{\sc tmcm-mul} and \mbox{\sc tmcm-crk} architectures. The SAT-based~\cite{subramanyan15} and query attacks were run on these obfuscated TMCM blocks, where the time limit was set to 2 days. Fig.~\ref{fig:attacks_random} presents the run-time of these attacks.

Observe from Fig.~\ref{fig:attacks_random} that as $n$ and $p$ increase, the \mbox{run-time} of the query attack increases slightly. Note that while the query attack can find the secret key of each instance, the \mbox{SAT-based} attack can find a solution on 39 and 43 instances under the \mbox{\sc tmcm-mul} and \mbox{\sc tmcm-crk} architectures, respectively. Observe that the query attack runs faster than the SAT-based attack on these instances. Note that Section~\ref{sec:results} presents more results of the query attack on different multiplier blocks obfuscated and locked by different techniques.

	\section{Proposed Hybrid Protection Technique}
\label{sec:defense}

This section initially presents the obfuscation technique used to hide filter coefficients behind decoys in the CAVM and MCM blocks of parallel direct and transposed forms of FIR filters (cf. Section \ref{subsec:obf_cavm} and Section \ref{subsec:obf_mcm},  respectively). Then, it describes the logic locking method using a point function described at RTL (cf. Section \ref{subsec:okpf}). Finally, it introduces the hybrid protection technique including both of these methods (cf. Section \ref{subsec:hybrid_protection}).

\begin{algorithm}[t]
	\small
	\caption{Selection of decoys for original constants}
	\begin{algorithmic}[1]
		\Statex \textbf{Inputs:} Original constants $C=\{c_1, c_2, \ldots, c_{n}\}$ and $v$ key bits.
		\Statex \textbf{Output:} Decoy set $D$.
		\State $noi = 0$ \Comment{Number of iterations}
		\State $nok = 0$ \Comment{Number of used key bits}
		\State $D = \emptyset$ \Comment{Set of $n$ decoy constant arrays}
		\While{$nok < v$}
		\State $nod = 2^{noi}$ \Comment{Number of decoys to be assigned}
		\For{$i=1$ \textbf{to}  $n$}
		\State $D_i$ = AssignDecoy($D_i$, $c_{i}$, $nod$)
		\State $nok = nok + 1$
		\If{$nok == v$}
		\State \textbf{break}
		\EndIf
		\EndFor
		\State $noi = noi + 1$
		\EndWhile
	\end{algorithmic}
	\label{algo:decoys}
\end{algorithm}

The original constants can be obfuscated using decoys as described in~\cite{aksoy21}. The motivation behind such obfuscation is that the use of decoys enables us to control the tradeoff between hardware complexity, output corruption, and filter behavior~\cite{aksoy21, aksoy22} when compared to logic locking. The obfuscation technique using decoys requires two main steps: i)~given the number of key bits, determine decoys for each original constant; ii)~realize the obfuscated design, where original constants are hidden behind decoys using MUXes and key bits. The selection of decoys for the original constants is done as shown in Algorithm~\ref{algo:decoys}. In its \textit{AssignDecoy} function (line 7), decoy selection can be done based on a given criterion, namely hardware complexity, output corruption, and filter behavior. In these criteria, decoys are chosen to be unique to increase the obfuscation. 

\subsection{Hardware Obfuscation of the CAVM Block}
\label{subsec:obf_cavm}

\begin{figure}[t]
	\centering
	\vspace*{-2mm}
	\includegraphics[width=8.5cm]{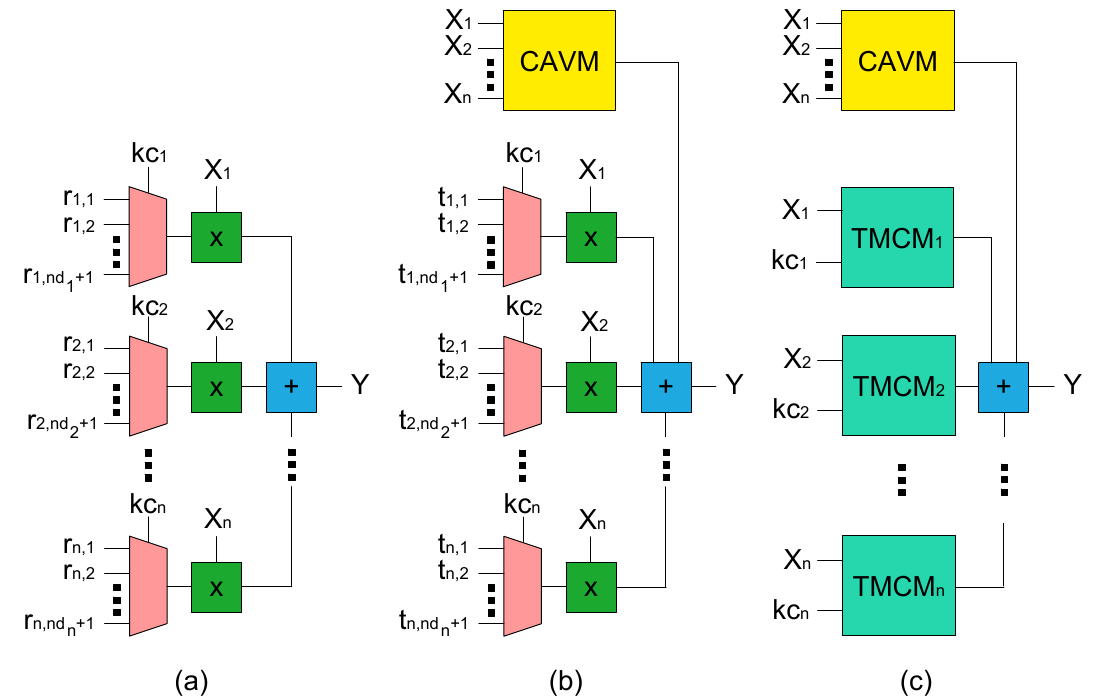}
	\vspace*{-4mm}
	\caption{Realizations of the obfuscation of the CAVM block using decoys: (a)~straightforward design; (b)~{\sc cavm-mul}; (c)~{\sc cavm-sa}.}
	\label{fig:obf_cavm}
	\vspace*{-4mm}
\end{figure}

Given $1 \times n$ original constant array $C = [c_1, c_2, \ldots, c_{n}]$ and the number of key bits for obfuscation, i.e., $v$, let $D$ denote a set of $n$ decoy constant arrays, i.e., $D = \{ [d_{1}^{1}, \ldots, d_{1}^{nd_1}], [d_{2}^{1}, \ldots, d_{2}^{nd_2}], \ldots, [d_{n}^{1}, \ldots, d_{n}^{nd_{n}}] \}$, where $nd_i$ is the number of decoy constants selected for the $i^{th}$ original constant determined based on a given criterion with $1 \leq i \leq n$. Then, the set $R$, which includes each original constant and its decoys, i.e., $R_i = c_i \cup D_i = [ c_i, d_i^1, \ldots, d_i^{nd_i} ]$ with $1 \leq i \leq n$, is formed. Let $r_{i,j}$ denote the $j^{th}$ constant in $R_i$ with $1 \leq i \leq n$ and $1 \leq j \leq nd_i+1$. Thus, the straightforward realization of the obfuscated CAVM block is given in Fig.~\ref{fig:obf_cavm}(a). Note that the key bits determined for each constant, i.e., $kc_i$, have the size of $\lceil log_2(nd_i+1) \rceil$ with $1 \leq i \leq n$. The secret key, which is formed as the concatenation of these key bits, is determined based on the location of the original constant in the constant array $R_i$.

\begin{figure}[t]
	\centering
	\includegraphics[width=9.0cm]{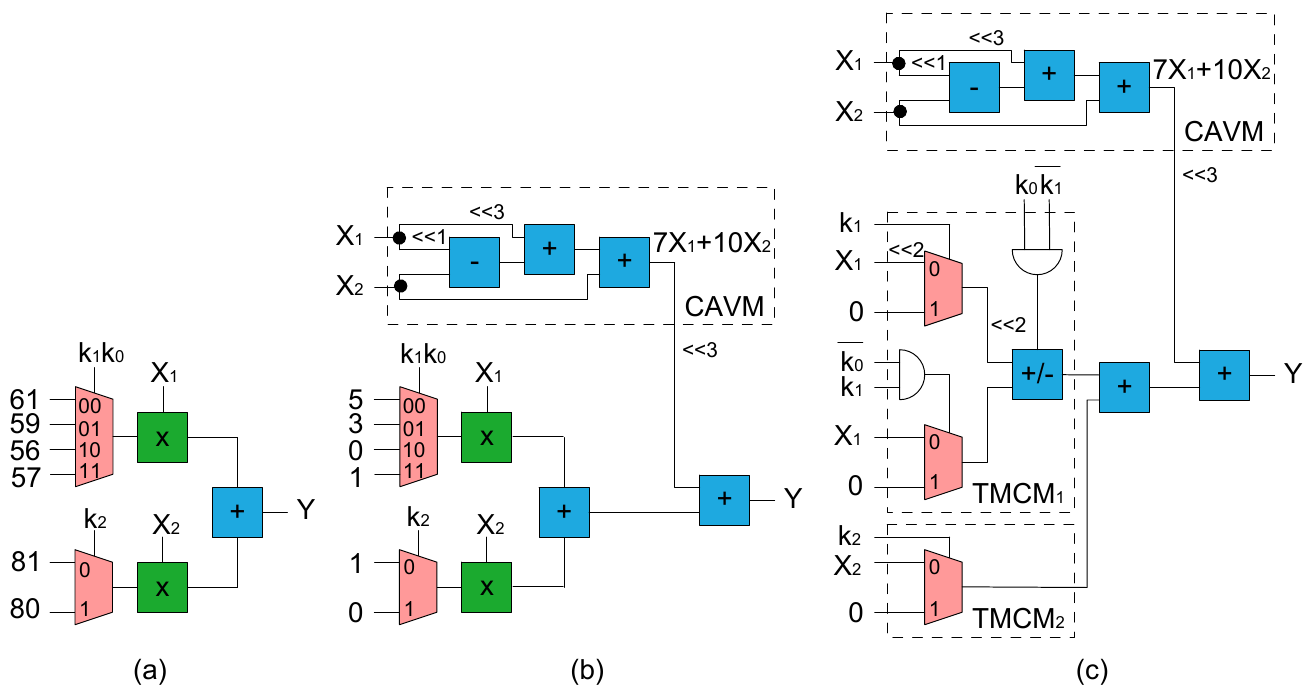}
	\vspace*{-8mm}
	\caption{Realizations of the obfuscation of the CAVM block including constants 57 and 81: (a)~straightforward design; (b)~{\sc cavm-mul}; (c)~{\sc cavm-sa}.}
	\label{fig:obf_cavm_ex}
	\vspace*{-6mm}
\end{figure}

Note that the size of a multiplier given in Fig.~\ref{fig:obf_cavm}(a) is related to the bit-width of the original constant and its decoy(s). Hence, to reduce the hardware complexity of the straightforward design, the size of constants, which are inputs of MUXes, can be decreased. To do so, we implement a CAVM block, where each entry of its constant array $S$ is an element of each $R$ array, i.e., $S = [s_1, s_2, \dots, s_n]$ with $s_i \in R_i = [c_i, d_i^1,\ldots, d_i^{nd_i} ]$ and $1 \leq i \leq n$. Then, the original constant and its decoys at inputs of each MUX are computed as $T_i = R_i - s_i$ with $1 \leq i \leq n$. Fig.~\ref{fig:obf_cavm}(b) presents the obfuscated design under the proposed architecture called {\sc cavm-mul}. Note that the CAVM block realizes $s_1X_1 + s_2X_2 + \ldots + s_nX_n$ and is implemented under the shift-adds architecture using the algorithm of~\cite{aksoy14_cavm}. The constants to be in $S$ are decided based on the hardware complexity of the CAVM block and the size of multipliers. This problem is formulated as a 0-1 Integer Linear Programming (ILP) problem. 

To further reduce the hardware complexity of the design in Fig.~\ref{fig:obf_cavm}(b), each multiplier with a MUX, which represents a TMCM block, is realized under the shift-adds architecture using the algorithm of~\cite{aksoy14_tmcm}. Fig.~\ref{fig:obf_cavm}(c) presents the obfuscated design under the proposed architecture called {\sc cavm-sa}.

Returning to our example in Fig.~\ref{fig:cavm} with $C = [57, 81]$ and assuming that the number of key bits is 3, the set $D$, that includes decoys for each constant, is found as $D=\{[61, 59, 56], [80]\}$ based on the hardware complexity criterion. Thus, the set $R$ is formed as $R = \{[61, 59, 56, 57], [81, 80]\}$. The straightforward realization of the obfuscated CAVM block using decoys is shown in Fig.~\ref{fig:obf_cavm_ex}(a), where the secret key is $\bm{K} = k_2k_1k_0 = 011$. Under the {\sc cavm-mul} and \mbox{{\sc cavm-sa}} architectures, the constant array $S$ is determined as $S = [56, 80]$. Thus, the set $T$ is formed as $\{[5, 3, 0, 1], [1, 0]\}$ leading to multipliers with smaller sizes when compared to those given in Fig.~\ref{fig:obf_cavm_ex}(a). The realization of the obfuscated CAVM design under the {\sc cavm-mul} architecture is given in Fig.~\ref{fig:obf_cavm_ex}(b). Furthermore, Fig.~\ref{fig:obf_cavm_ex}(c) presents the shift-adds realization of the TMCM blocks implementing the constant multiplications including those in the set $T$ under the \mbox{{\sc cavm-sa}} architecture.

\begin{figure}[t]
	\centering
	\includegraphics[width=8.5cm]{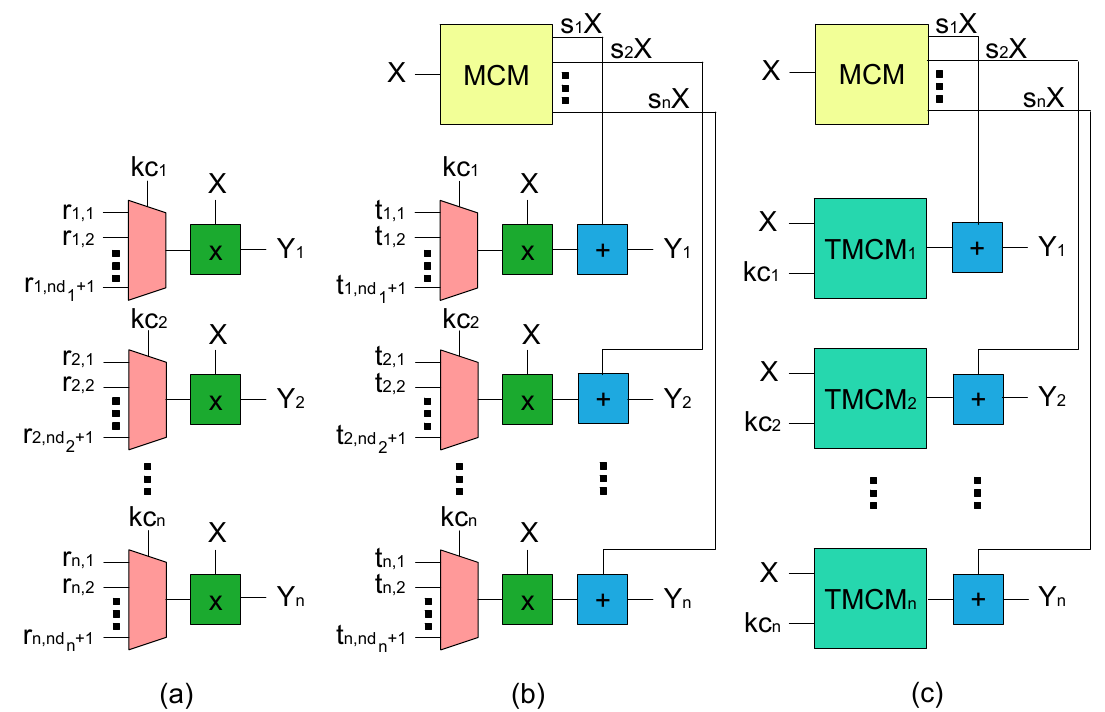}
	\vspace*{-2mm}
	\caption{Realizations of the obfuscation of the MCM block using decoys: (a)~straightforward design; (b)~{\sc mcm-mul}; (c)~{\sc mcm-sa}.}
	\label{fig:obf_mcm}
	\vspace*{-4mm}
\end{figure}

\begin{figure}[t]
	\centering
	\includegraphics[width=8.5cm]{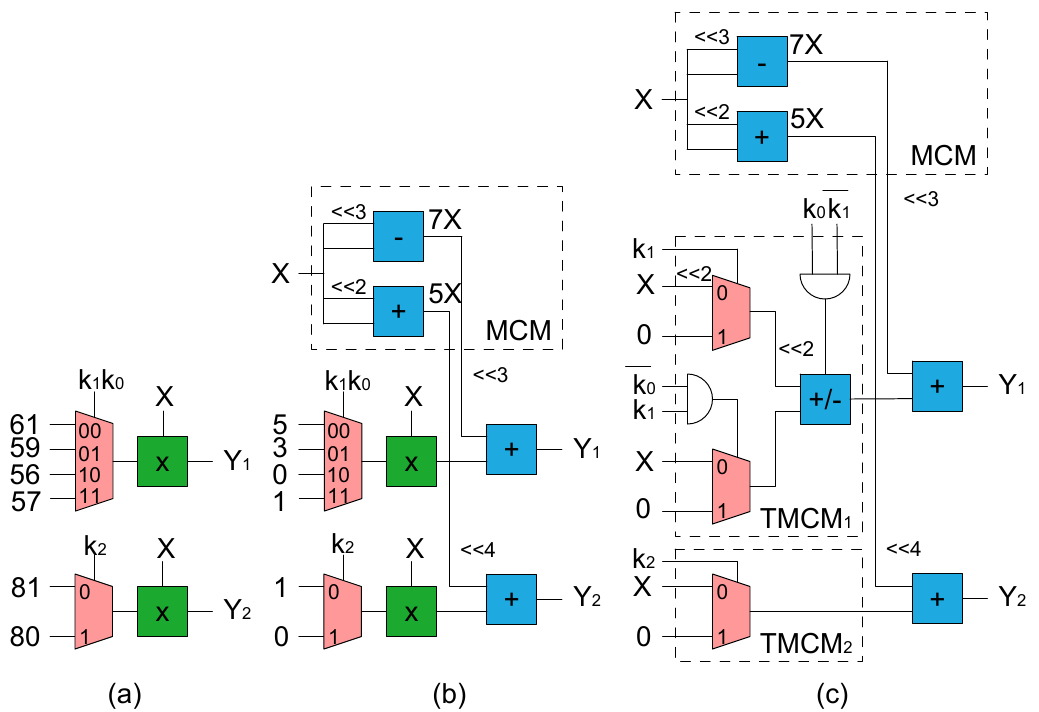}
	\vspace*{-2mm}
	\caption{Realizations of the obfuscation of the MCM block including constants 57 and 81: (a)~straightforward design; (b)~{\sc mcm-mul}; (c)~{\sc mcm-sa}.}
	\label{fig:obf_mcm_ex}
	\vspace*{-4mm}
\end{figure}

In addition to the obfuscation using decoys on the CAVM block, we also developed the constant obfuscation technique used in the ASSURE tool~\cite{pilato21}. Given the number of key bits, constants in the original CAVM block are replaced by key bits under the architecture called {\sc cavm-crk}.

\subsection{Hardware Obfuscation of the MCM Block}
\label{subsec:obf_mcm}

Similarly, the MCM block can also be obfuscated using decoys. After decoys selected for each original constant are found based on the given criterion, and the set $R$ is determined, the straightforward realization of the obfuscated design can be obtained as illustrated in Fig.~\ref{fig:obf_mcm}(a). Moreover, the size of multipliers can be reduced by determining the set of constants $S$ from the set $R$ and the set $T$ is computed accordingly as described in Section~\ref{subsec:obf_cavm}. The multiplications of constants in the set $S$ by the variable $X$, i.e., $s_1X, s_2X, \ldots, s_nX$, are realized in an MCM block, which is implemented under the shift-adds architecture using the algorithm of~\cite{aksoy10}. Fig.~\ref{fig:obf_mcm}(b) presents the obfuscated design under the proposed architecture called {\sc mcm-mul}. Furthermore, the multiplierless realization of the design obfuscated under the {\sc mcm-mul} architecture can be obtained by realizing the TMCM block under the \mbox{shift-adds} architecture using the algorithm of~\cite{aksoy14_tmcm}. Fig.~\ref{fig:obf_mcm}(c) shows the obfuscated design under the proposed architecture called {\sc mcm-sa}.

Returning to our example, Fig.~\ref{fig:obf_mcm_ex} presents the straightforward realization of the obfuscated MCM block and its designs under the {\sc mcm-mul} and {\sc mcm-sa} architectures.

In addition to the obfuscation using decoys, constants in the original MCM block are replaced by key bits under the architecture called {\sc mcm-crk}.

\begin{figure}[t]
	\centering
	\includegraphics[width=8.5cm]{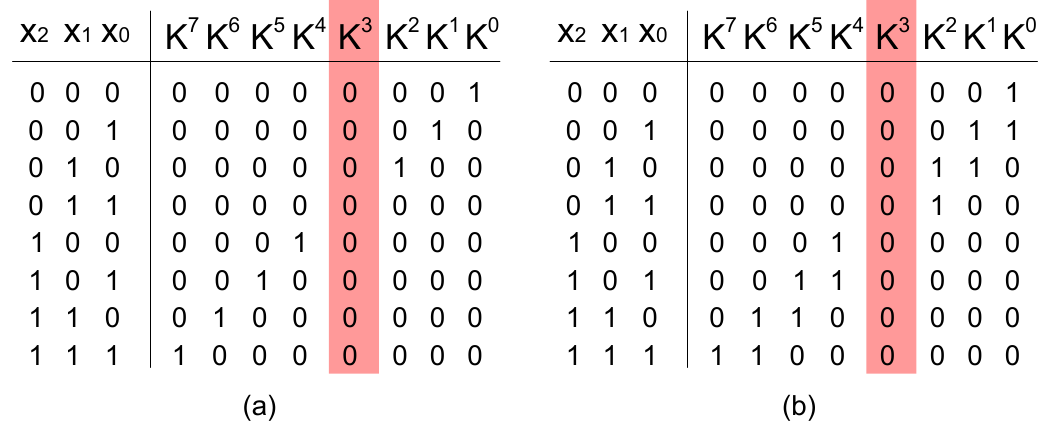}
	\vspace*{-2mm}
	\caption{(a)~Behavior of a Boolean function locked by one-point function; (b)~behavior of a Boolean function locked by relaxed one-point function.}
	\label{fig:pf}
	\vspace*{-6mm}
\end{figure}

\subsection{Logic Locking with a Point Function}
\label{subsec:okpf}

As shown in Sections~\ref{sec:attack} and~\ref{sec:results}, the constant obfuscation techniques are vulnerable to the SAT-based attack and its variants. The motivation behind locking the obfuscated design with a point function is to make it resilient to these techniques. In order to increase the number of DIPs to be explored in a SAT-based attack, one can lock primary outputs of a multiplier block using a point function\footnote{A one-point function is a Boolean function that evaluates to 1 at exactly one input pattern.} at RTL as done at gate-level in~\cite{yasin16, yasin17, xie19, shakya19}. 

Suppose that a Boolean function $f: \mathbb{B}^q \rightarrow \mathbb{B}$ is locked using a one-point function with $w$ key bits, where $w \leq q$, leading to a locked Boolean function $g: \mathbb{B}^{w+q} \rightarrow \mathbb{B}$ and let $\bm{K}$ denotes the secret key. Then, $f(X) = g(X,\bm{K})$ under all possible input values. Fig.~\ref{fig:pf}(a) shows the behavior of the locked function $g$ under each possible key value when $q=w=3$ and $k_2k_1k_0 = 011$ is the secret key. In this figure, $K^i$ stands for the assignment of the value $i$ in binary to key bits, i.e., $k_{w-1} \ldots k_1k_0 = (i)_{bin}$ with $0 \leq i \leq 2^w-1$. Also, the value of logic 0 (1) under each possible key value denotes that the locked function $g$ is (not) equal to the original function $f$. Note that the locked function under the secret key, i.e., $\bm{K} = K^3$ highlighted in our example, always generates the same output as the original function for every input pattern. Observe from Fig.~\ref{fig:pf}(a) that each input pattern eliminates at most one wrong key, leading to an exponential number of DIPs to find the secret key, i.e., $2^w$. Moreover, such a one-point function can be relaxed to increase the corruption at a primary output. For example, Fig.~\ref{fig:pf}(b) presents the behavior of the locked function $g$, where each input pattern can eliminate at most 2 wrong keys. Observe that the exponential number of tries to find the secret key is still valid, i.e., $2^{w-1}$ in this case. Furthermore, multiple primary outputs can be locked using point functions with different key bits.

\begin{lstlisting}[mathescape=true, float=t, language=verilog, caption=Logic locking using a point function at RTL., label=lst:pf_code]
// One-point function at RTL
always @(*) begin 
	if ($X$ == $K$) 
		if ($K$ == $\bm{K}$)
			g = f; 
		else 
			g = !f; 
	else
		g = f; 
end 
//Relaxed one-point function at RTL
always @(*) begin 
	if (($X$ - $K$) >= q'd0 && ($X$ - $K$) <= q'dcv)
		if ($K$ == $\bm{K}$)
			g = f; 
		else
			g = !f;
	else
		g = f; 
end 
\end{lstlisting}

\begin{figure}[t]
	\centering
	\vspace*{-2mm}
	\includegraphics[width=7.5cm]{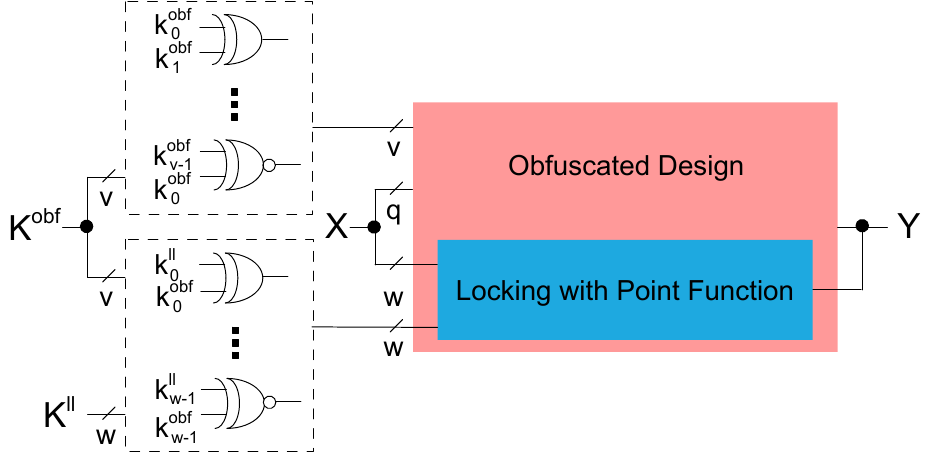}
	\vspace*{-2mm}
	\caption{Proposed hybrid protection technique.}
	\label{fig:pf_obf}
	\vspace*{-4mm}
\end{figure}

Listing~\ref{lst:pf_code} presents the Verilog code snippet, which describes logic locking using the one-point function at RTL. In this code, $X$ is an array of primary inputs and $K$ is an equally sized array of key bits. Moreover, the relaxed one-point function can also be described at RTL as shown in the same listing. In its code, $cv$ stands for the corruption value, which denotes the maximum number of wrong keys that can be distinguished by a single input pattern.

We note that since the point function is described at RTL, the synthesis tool shapes its circuit based on the given synthesis parameters. Thus, its realization does not have a regular structure like logic locking methods of~\cite{xie19, shakya19, nguyen21}. 



\subsection{Combination of Obfuscation and Logic Locking}
\label{subsec:hybrid_protection}

The hybrid protection technique includes obfuscation and logic locking using a point function described in previous subsections. Initially, the obfuscation technique using decoys is applied and then, the obfuscated design is locked using a point function. Fig.~\ref{fig:pf_obf} illustrates the hybrid protection technique, where $X$ and $Y$ denote the inputs and outputs of the original design and $K^{obf}$ and $K^{ll}$ stand for the key bits used for obfuscation and locking, respectively. Additionally, to thwart the structural attacks, the key bits used for logic locking, $K^{ll}$, are hidden among the key bits used for obfuscation, $K^{obf}$, using {\sc xor/xnor} gates. In this scheme, an {\sc xor} ({\sc xnor}) gate, which has $k_{i}^{ll}$ and $k_j^{obf}$ as inputs, is generated if the value of $k_j^{obf}$ in the secret key is equal to logic 0 (1) value, where \mbox{$0 \leq i \leq w-1$} and $0 \leq j \leq v-1$. Then, the output of this gate is connected to the net, which would be driven by $k_i^{ll}$. Moreover, to thwart the query attack, each $k_i^{obf}$ is hidden among another $k_j^{obf}$ using an {\sc xor/xnor} gate, where $i \neq j$ and $0 \leq i,j \leq v-1$. By doing so, each key bit is observed with other key bits at a primary output, making it harder for the query attack to prove the value of the related key bit.

We developed a Computer-Aided Design (CAD) tool to automate the design and verification process for the obfuscation and locking of the CAVM, MCM, and TMCM blocks and the parallel direct and transposed form and folded FIR filters. The CAD tool takes the filter coefficients, the number of key bits, the design architecture, and other design parameters as inputs and generates the description of the obfuscated design in Verilog, the testbench for verification, and synthesis and simulation scripts. Note that designs are described in a behavioral fashion at RTL. 

	\section{Experimental Results}
\label{sec:results}

In this section, we introduce the gate-level synthesis results of multiplier blocks protected by the proposed hybrid technique, obfuscated by previously proposed techniques, and locked by prominent logic locking methods. We also provide the results of well-known logic locking attacks and the proposed query attack on these designs. Furthermore, we present the results of obfuscated and locked FIR filters and also, introduce the results of prominent attacks on these designs. Finally, we explore the impact of parameters used in the point function on the hardware complexity and resiliency to the \mbox{SAT-based} attack and present gate-level synthesis results of the obfuscated and locked CAVM block of the direct form filter, which has promising security properties. 

\begin{table}[t]
	\centering
	\caption{Details on the fir filter and its multiplier blocks.} 
	\vspace*{-2mm}
	\begin{tabular}{|c|cc|c|cc|}
		\hline
		Filter & $n$ & $mbw$ & Multiplier Block & $\#in$ & $\#out$ \\
		\hline
		\multirow{3}{*}{Johansson08} & \multirow{3}{*}{30}  & \multirow{3}{*}{10} & CAVM & 480  & 31 \\
		&                      &                     & MCM  & 16   & 733 \\
		&                      &                     & TMCM & 21   & 26 \\
		\hline
	\end{tabular}
	\label{tab:firs}
	\vspace*{-4mm}
\end{table}

\begin{table*}[t]
	\centering
	\caption{Results of obfuscated and protected multiplier blocks.} 
	\vspace*{-2mm}
	\begin{tabular}{|l|l|l|ccc|cccccc|cc|}
		\hline
		\multicolumn{1}{|c|}{ \multirow{3}{*}{Block} }  & \multicolumn{1}{c|}{ \multirow{3}{*}{Architecture} } & \multicolumn{1}{c|}{ \multirow{3}{*}{Technique} } & \multicolumn{3}{c|}{ \multirow{2}{*}{Synthesis} } & \multicolumn{8}{c|}{Attacks} \\
		\cline{7-14}
		&                                               &                                                   &      &       &                                      & SAT & ATPG & AppSAT & DoubleDIP & \multicolumn{2}{c|}{Query} & \multicolumn{2}{c|}{SCOPE} \\
		\cline{4-6} \cline{7-14}
		&                                               &                                                   & area & delay & power                                & time & time & time & time & prv & time & cdk/dk & time \\
		\hline 
		&                                                                                                   & Decoy~\cite{aksoy21}  & 15435 & 4616 & 4641 & 155143 & OoT  & OoT & OoT & 32 & 9893  & 20/32 & 8 \\ 
		& \multirow{-2}{*}{\sc cavm-mul}                        & \ccsb{Proposed Hybrid} & \ccsb{15710} & \ccsb{4611} & \ccsb{4757} & \ccsb{OoT}   & \ccsb{OoT}  & \ccsb{OoT} & \ccsb{OoT} & \ccsb{0}  & \ccsb{OoT}   & \ccsb{1/1}   & \ccsb{13} \\  
		\cline{2-14}
		&                                                                                                   & Decoy~\cite{aksoy21}  & 15465 & 4611 & 4475 & 36083 & 4539 & OoT & OoT & 32 & 9944  & 20/32 & 8 \\  
		& \multirow{-2}{*}{\sc cavm-sa}                         & \ccsb{Proposed Hybrid} & \ccsb{15704} & \ccsb{4715}  & \ccsb{4497} & \ccsb{OoT}   & \ccsb{OoT}  & \ccsb{OoT} & \ccsb{OoT} & \ccsb{0}  & \ccsb{29328} & \ccsb{1/1}   & \ccsb{12} \\ 
		\cline{2-14}
		&                                                                                                   & Constant~\cite{pilato21} & 18737 & 3982 & 4756 & 110   & 1446 & OoT & OoT & 32 & 897   & 21/27 & 11 \\ 
		\multirow{-6}{*}{CAVM} & \multirow{-2}{*}{\sc cavm-crk} & \ccsb{Proposed Hybrid}  & \ccsb{18976}  & \ccsb{4265} & \ccsb{4809} & \ccsb{OoT}   & \ccsb{OoT}  & \ccsb{OoT} & \ccsb{OoT} & \ccsb{0}  & \ccsb{1937}  & \ccsb{2/3}   & \ccsb{16} \\ 
		\hline
		&                                                                                                   & Decoy~\cite{aksoy21}  & 10949 & 3102 & 2839 & 106 & OoT & 324 & 243 & 32 & 176 & 27/32 & 7  \\
		& \multirow{-2}{*}{\sc mcm-mul}                         & \ccsb{Proposed Hybrid} & \ccsb{11173} & \ccsb{3031} & \ccsb{2897} & \ccsb{OoT} & \ccsb{OoT} & \ccsb{OoT} & \ccsb{OoT} & \ccsb{0}  & \ccsb{197} & \ccsb{1/1}   & \ccsb{11} \\
		\cline{2-14}
		&                                                                                                   & Decoy~\cite{aksoy21}  & 10493 & 3112 & 2493 & 119 & OoT & 342 & 254 & 32 & 152 & 27/32 & 7  \\
		& \multirow{-2}{*}{\sc mcm-sa}                          & \ccsb{Proposed Hybrid} & \ccsb{10705} & \ccsb{3159} & \ccsb{2495} & \ccsb{OoT} & \ccsb{OoT} & \ccsb{OoT} & \ccsb{OoT} & \ccsb{0}  & \ccsb{115} & \ccsb{1/1}   & \ccsb{11} \\
		\cline{2-14}
		&                                                                                                   & Constant~\cite{pilato21} & 12799 & 2772 & 2412 & 159 & OoT & 415 & 300 & 32 & 459 & 18/32 & 7  \\
		\multirow{-6}{*}{MCM}  & \multirow{-2}{*}{\sc mcm-crk}  & \ccsb{Proposed Hybrid}   & \ccsb{13038} & \ccsb{2970} & \ccsb{2456} & \ccsb{OoT} & \ccsb{OoT} & \ccsb{OoT} & \ccsb{OoT} & \ccsb{0}  & \ccsb{759} & \ccsb{1/1}   & \ccsb{11} \\
		\hline
		&                                                                                                   & Decoy~\cite{aksoy21}  & 1545 & 3517 & 610  & 241  & 6783  & 378  & 496 & 32 & 7  & 21/32 & 4 \\
		& \multirow{-2}{*}{\sc tmcm-mul}                        & \ccsb{Proposed Hybrid} & \ccsb{1794} & \ccsb{4278} & \ccsb{639}  & \ccsb{OoT}  & \ccsb{OoT}   & \ccsb{OoT}  & \ccsb{OoT} & \ccsb{0}  & \ccsb{17} & \ccsb{2/3}   & \ccsb{2} \\
		\cline{2-14}
		&                                                                                                   & Decoy~\cite{aksoy21}  & 2043 & 4452 & 1037 & 738  & 963   & 935  & OoT & 32 & 37 & 17/32 & 4 \\
		& \multirow{-2}{*}{\sc tmcm-sa}                         & \ccsb{Proposed Hybrid} & \ccsb{2245} & \ccsb{4536} & \ccsb{1065} & \ccsb{OoT}  & \ccsb{OoT}   & \ccsb{OoT}  & \ccsb{OoT} & \ccsb{0}  & \ccsb{42} & \ccsb{1/2}   & \ccsb{3} \\
		\cline{2-14}
		&                                                                                                   & Constant~\cite{pilato21} & 1566 & 2997 & 623  & 1035 & 15571 & 1032 & OoT & 32 & 29 & 20/32 & 4 \\
		\multirow{-6}{*}{TMCM} & \multirow{-2}{*}{\sc tmcm-crk} & \ccsb{Proposed Hybrid}   & \ccsb{1776} & \ccsb{3655} & \ccsb{642}  & \ccsb{OoT}  & \ccsb{OoT}   & \ccsb{OoT}  & \ccsb{OoT} & \ccsb{0}  & \ccsb{48} & \ccsb{1/1}   & \ccsb{2} \\
		\hline
	\end{tabular}
	\label{tab:obf_results}
	\vspace*{-4mm}
\end{table*}

\begin{table*}[t]
	\centering
	\caption{Results of locked multiplier blocks.} 
	\vspace*{-2mm}
	\begin{tabular}{|l|l|ccc|cccccc|cc|}
		\hline
		\multicolumn{1}{|c|}{ \multirow{3}{*}{Block} }  & \multicolumn{1}{c|}{ \multirow{3}{*}{Logic Locking} } & \multicolumn{3}{c|}{ \multirow{2}{*}{Synthesis} } & \multicolumn{8}{c|}{Attacks} \\
		\cline{6-13}
		&                                                 & & &                                                          & SAT & ATPG & AppSAT & DoubleDIP & \multicolumn{2}{c|}{Query} & \multicolumn{2}{c|}{SCOPE} \\
		\cline{3-13}
		&                                                 & area & delay & power                                         & time & time & time & time & prv & time & cdk/dk & time \\
		\hline 
		& RLL                  & 16411 & 3385 & 4183 & 171 & OoT & 205   & 2609 & 32 & 254 & 0/0 & 9 \\ 
		& \cccb{RLL+AntiSAT}   & \cccb{16492} & \cccb{3416} & \cccb{4127} & \cccb{OoT} & \cccb{OoT} & \cccb{OoT}   & \cccb{OoT}  & \cccb{16} & \cccb{364} & \cccb{0/0} & \cccb{14} \\  
		& \cccb{RLL+CASLock}   & \cccb{16473} & \cccb{3502} & \cccb{4110} & \cccb{OoT} & \cccb{OoT} & \cccb{OoT}   & \cccb{OoT}  & \cccb{16} & \cccb{443} & \cccb{0/0} & \cccb{13} \\  
		& RLL+SARLock          & 16512 & 3572 & 4191 & OoT & OoT & 48855 & OoT  & 32 & 441 & 8/8 & 13\\ 
		& RLL+SFLL             & 16506 & 3473 & 4205 & 339 & 829 & 39664 & OoT  & 32 & 436 & 0/0 & 14\\ 
		\multirow{-6}{*}{CAVM} & \cccb{RLL+SKGLock} & \cccb{16559} & \cccb{3894} & \cccb{4308} & \cccb{OoT} & \cccb{OoT} & \cccb{OoT}   & \cccb{OoT}  & \cccb{19} & \cccb{513} & \cccb{5/6} & \cccb{14}\\ 
		\hline
		& RLL                  & 8090 & 2398 & 2039 & 86  & 122  & 293   & 371 & 32 & 89  & 0/0   & 6\\ 
		& \cccb{RLL+AntiSAT}   & \cccb{8244} & \cccb{2434} & \cccb{2065} & \cccb{OoT} & \cccb{OoT}  & \cccb{OoT}   & \cccb{OoT} & \cccb{16} & \cccb{249} & \cccb{0/0}   & \cccb{9} \\  
		& RLL+CASLock          & 8121 & 2404 & 2024 & OoT & OoT  & 1054  & OoT & 16 & 164 & 0/0   & 9\\  
		& RLL+SARLock          & 8209 & 2467 & 2041 & OoT & OoT  & 45013 & OoT & 32 & 228 & 10/10 & 9\\ 
		& RLL+SFLL             & 8166 & 2452 & 2019 & 576 & 7870 & 2626  & OoT & 32 & 243 & 0/0   & 9\\ 
		\multirow{-6}{*}{MCM}  & \cccb{RLL+SKGLock} & \cccb{8252} & \cccb{2464} & \cccb{2056} & \cccb{OoT} & \cccb{OoT}  & \cccb{OoT}   & \cccb{OoT} & \cccb{19} & \cccb{160} & \cccb{5/5}   & \cccb{9}\\ 
		\hline
		& RLL                  & 1587 & 3712 & 646 & 8    & 39   & 57   & 77  & 32 & 15 & 0/0   & 2\\ 
		& \cccb{RLL+AntiSAT}   & \cccb{1659} & \cccb{3545} & \cccb{632} & \cccb{OoT}  & \cccb{OoT}  & \cccb{OoT}  & \cccb{OoT} & \cccb{13} & \cccb{28} & \cccb{0/0}   & \cccb{2}\\  
		& \cccb{RLL+CASLock}   & \cccb{1653} & \cccb{3664} & \cccb{628} & \cccb{OoT}  & \cccb{OoT}  & \cccb{OoT}  & \cccb{OoT} & \cccb{15} & \cccb{20} & \cccb{0/0}   & \cccb{2} \\  
		& RLL+SARLock          & 1659 & 3756 & 658 & OoT  & OoT  & 2662 & OoT & 31 & 36 & 6/6   & 3\\ 
		& RLL+SFLL             & 1644 & 3800 & 653 & 1095 & 1089 & 5363 & OoT & 32 & 36 & 0/0   & 2\\ 
		\multirow{-6}{*}{TMCM} & \cccb{RLL+SKGLock} & \cccb{1694} & \cccb{3739} & \cccb{676} & \cccb{OoT}  & \cccb{OoT}  & \cccb{OoT}  & \cccb{OoT} & \cccb{18} & \cccb{34} & \cccb{10/10} & \cccb{2}\\ 
		\hline
	\end{tabular}
	\label{tab:lock_results}
	\vspace*{-4mm}
\end{table*}

\subsection{Results of Obfuscated and Locked Multiplier Blocks}

Based on our experimental observations, similar outcomes have been observed on FIR filters with a different number of coefficients and different bit-width of filter input and coefficients in comparison of design architectures, obfuscation and locking techniques, and attacks. Hence, in this experiment, a single FIR filter with a small number of coefficients and a small bit-width of filter input and coefficients was used to reveal the effectiveness of obfuscation and locking techniques. Table~\ref{tab:firs} presents the details of this filter taken from~\cite{firsuite}, where \textit{\#in} and \textit{\#out} are respectively the number of inputs and outputs of the multiplier blocks when $ibw$ is 16.

Table~\ref{tab:obf_results} presents the synthesis results of the CAVM, MCM, and TMCM blocks of the FIR filter obfuscated by previously proposed methods~\cite{aksoy21, pilato21} and protected by the proposed hybrid technique. Note that the {\sc tmcm-sa} architecture denotes the TMCM block obfuscated using decoys under \mbox{the shift-adds} architecture. Logic synthesis was performed by Cadence Genus using a commercial 65\;nm cell library with the aim of area optimization. For this aim, a very high virtual clock period value, i.e., 80\;ns, was used. The encrypted designs were validated by simulation using 10,000 randomly generated inputs, where the switching activity data of each node in the design were collected and stored in a Switching Activity Interchange Format (SAIF) file, which is later used by the synthesis tool while computing the power dissipation. For obfuscation, 32 key bits were used. There were 16 key bits for locking using the one-point function. Thus, a total of 48 key bits were used in designs protected by the hybrid technique. In this table, \textit{area}, \textit{delay}, and \textit{power} stand for the total area in $\mu m^2$, delay in the critical path in $ps$, and total power dissipation in $\mu W$, respectively. This table also presents the results of OG attacks, namely SAT- and ATPG-based attacks, the approximate AppSAT attack taken from~\cite{host15}, and the DoubleDIP attack taken from~\cite{doubledip}, and the OL SCOPE attack taken from~\cite{scope}. For the SCOPE attack, \textit{cdk} and \textit{dk} denote the number of correctly deciphered key bits and the number of deciphered key bits, respectively. The \mbox{time limit} given to the attacks was 2 days. In this table, designs, whose secret key has not been discovered by the given attacks, are highlighted.

\begin{table*}[t]
	\centering
	\caption{Results of Obfuscated and Locked FIR Filters.} 
	\vspace*{-2mm}
	\begin{tabular}{|@{\hskip3pt}l@{\hskip3pt}|l@{\hskip3pt}|l@{\hskip3pt}|c@{\hskip3pt}c@{\hskip3pt}c@{\hskip3pt}|c|@{\hskip3pt}c@{\hskip3pt}c@{\hskip3pt}|l@{\hskip3pt}|c@{\hskip3pt}c@{\hskip3pt}c@{\hskip3pt}|c|@{\hskip3pt}c@{\hskip3pt}c@{\hskip3pt}|}
		\hline
		\multicolumn{1}{|c|}{ \multirow{4}{*}{Filter} }    & \multicolumn{8}{c|}{ Obfuscation and Hybrid Protection }                                                                                                                                                                           & \multicolumn{7}{c|}{ Logic Locking } \\
		\cline{2-16}
		& \multicolumn{1}{c|}{ \multirow{3}{*}{Architecture} } & \multicolumn{1}{c|}{ \multirow{3}{*}{Technique} }   & \multicolumn{3}{c|}{ \multirow{2}{*}{Synthesis} }            & \multicolumn{3}{c|}{Attacks}     & \multicolumn{1}{c|}{ \multirow{3}{*}{Technique} } & \multicolumn{3}{c|}{ \multirow{2}{*}{Synthesis} } & \multicolumn{3}{c|}{Attacks} \\
		\cline{7-9} \cline{14-16}
		&                                                      &                                                     &      &       &                                               & KC2 & \multicolumn{2}{c|}{SCOPE} &                                                 &      &       &                                      & KC2 & \multicolumn{2}{c|}{SCOPE} \\
		\cline{4-9} \cline{11-16}
		&                                                      &                                                     & area & delay & power                                         & time & cdk/dk & time             &                                                 & area & delay & power                                & time & cdk/dk & time  \\
		\hline
		&                                                      & Decoy~\cite{aksoy21}     & 19238 & 4907 & 3088 & Failed & 18/32 & 8   & RLL         & 20233 & 3566 & 2906 & Failed & 15/18 & 10  \\
		&\multirow{-2}{*}{\sc cavm-mul}                        & Proposed Hybrid   & 19500 & 4828 & 3153 & OoT    & 1/1   & 13  & RLL+AntiSAT & 20300 & 3512 & 2896 & Failed & 21/26 & 14  \\
		\cline{2-9}                                                                                                                                                                  
		&                                                      & Decoy~\cite{aksoy21}     & 19243 & 4792 & 3012 & Failed & 19/32 & 8   & RLL+CASLock & 20324 & 3746 & 2928 & OoT    & 11/15 & 14  \\
		&\multirow{-2}{*}{\sc cavm-sa}                         & Proposed Hybrid   & 19485 & 4798 & 3040 & OoT    & 1/1   & 13  & RLL+SARLock & 20326 & 3630 & 2936 & OoT    & 31/37 & 14  \\
		\cline{2-9}                                                                                                                                                            
		&                                                      & Constant~\cite{pilato21} & 22551 & 4228 & 2734 & Failed & 14/32 & 10  & RLL+SFLL    & 20307 & 3619 & 2892 & Failed & 21/28 & 14  \\
		\multirow{-6}{*}{Direct}                               & \multirow{-2}{*}{\sc cavm-crk}                        & Proposed Hybrid   & 22796 & 4244 & 2757 & OoT    & 2/4   & 15  & RLL+SKGLock & 20380 & 4052 & 2994 & Failed & 16/24 & 15  \\
		\hline                                                                                                                                                                   
		&                                                      & Decoy~\cite{aksoy21}     & 25195 & 3470 & 3848 & 100347 & 25/32 & 11  & RLL         & 22362 & 3093 & 3303 & 67811  & 16/21 & 9   \\
		&\multirow{-2}{*}{\sc mcm-mul}                         & Proposed Hybrid   & 25439 & 3540 & 3896 & OoT    & 1/1   & 16  & RLL+AntiSAT & 22510 & 3218 & 3302 & OoT    & 15/24 & 15  \\
		\cline{2-9}                                                                                                                                                                  
		&                                                      & Decoy~\cite{aksoy21}     & 24967 & 3322 & 3569 & 82952  & 26/32 & 10  & RLL+CASLock & 22461 & 3337 & 3320 & OoT    & 7/8   & 14  \\
		&\multirow{-2}{*}{\sc mcm-sa}                          & Proposed Hybrid   & 25139 & 3346 & 3562 & OoT    & 1/1   & 15  & RLL+SARLock & 22425 & 3183 & 3303 & OoT    & 31/41 & 14  \\
		\cline{2-9}                                                                                                                                                                 
		&                                                      & Constant~\cite{pilato21} & 27126 & 3240 & 3273 & 51973 & 21/32 & 11   & RLL+SFLL    & 22389 & 3116 & 3311 & OoT    & 19/29 & 14  \\
		\multirow{-6}{*}{Trans.}                           & \multirow{-2}{*}{\sc mcm-crk}                         & Proposed Hybrid   & 27433 & 3256 & 3290 & OoT   & 1/1   & 17   & RLL+SKGLock & 22514 & 3186 & 3329 & OoT    & 17/24 & 15  \\
		\hline                                                                                                                                                                     
		&                                                      & Decoy~\cite{aksoy21}     & 9126 & 4785 & 869  & 7478  & 20/32 & 2     & RLL         & 9183  & 4496 & 904 & 7845    & 15/18 & 3   \\
		&\multirow{-2}{*}{\sc tmcm-mul}                        & Proposed Hybrid   & 9379 & 4665 & 933  & OoT   & 2/2   & 3     & RLL+AntiSAT & 9225  & 4681 & 882 & OoT     & 8/15  & 4   \\                                                                                                                                      
		\cline{2-9}                                                                                                                                                                    
		&                                                      & Decoy~\cite{aksoy21}     & 9791 & 5758 & 1168 & 11895 & 18/32 & 2     & RLL+CASLock & 9237  & 4675 & 926 & OoT     & 8/13  & 4   \\
		&\multirow{-2}{*}{\sc tmcm-sa}                         & Proposed Hybrid   & 9966 & 5646 & 1236 & OoT   & 9/13  & 3     & RLL+SARLock & 9235  & 4761 & 911 & OoT     & 31/31 & 4   \\
		\cline{2-9}                                                                                                                                                                       
		&                                                      & Constant~\cite{pilato21} & 9126 & 4328 & 870  & 5657  & 22/32 & 2     & RLL+SFLL    & 9222  & 4570 & 892 & OoT     & 16/20 & 4   \\
		\multirow{-6}{*}{Folded}                               & \multirow{-2}{*}{\sc tmcm-crk}                        & Proposed Hybrid   & 9356 & 4602 & 894  & OoT   & 1/1   & 3     & RLL+SKGLock & 9288  & 4434 & 910 & OoT     & 18/31 & 4   \\
		\hline
	\end{tabular}
	\label{tab:fir_results}
	\vspace*{-4mm}
\end{table*}

\subsubsection{Comments on Hardware Complexity}

Observe from Table~\ref{tab:obf_results} that the hybrid protection technique increases the hardware complexity when compared to the obfuscation techniques simply due to the inclusion of the point function and logic for the obfuscation of key bits. Note that the increase of area in the CAVM, MCM, and TMCM blocks reaches up to 1.7\%, 2\%, and 13.8\%, respectively. The obfuscation and hybrid protection of the CAVM and MCM blocks under the proposed architectures, i.e., {\sc cavm-mul}, {\sc cavm-sa}, \mbox{\sc mcm-mul}, and {\sc mcm-sa}, lead to designs with less area when compared to those realized under the \mbox{{\sc cavm-crk}} and {\sc mcm-crk} architectures. Note that such a decrease reaches up to 17.6\% and 18\% in the CAVM and MCM blocks, respectively. This is simply because the proposed techniques exploit common subexpressions shared in constant multiplications. On the other hand, the obfuscation and hybrid protection of the TMCM blocks under the \mbox{\sc tmcm-mul} and {\sc tmcm-crk} architectures lead to designs with less area with respect to those realized under the {\sc tmcm-sa} architecture. Note that such a decrease reaches up to 24.3\%. This is because a single multiplier is replaced by a large number of addition and subtraction operations under the {\sc tmcm-sa} architecture. It is also observed that the minimum achievable delay values in the critical path of obfuscated and protected multiplier blocks are very close to each other, meaning that the inclusion of the point function and logic for the obfuscation of key bits does not have a significant impact while realizing the design with the smallest delay.

\subsubsection{Comments on Attack Resiliency}

Observe also from Table~\ref{tab:obf_results} that while the OG attacks can easily discover the secret key on the obfuscated designs, the OL attack can decipher all the key bits with high accuracy, except for the CAVM design obfuscated under the {\sc cavm-crk} architecture. On the other hand, none of these attacks can break the defense built by the hybrid protection technique. Note that the designs protected by the hybrid technique were also applied to \mbox{Fa-SAT}~\cite{limaye21} and the Valkyrie tool~\cite{limaye22}, but without any success due to the combination of both obfuscation and locking.

Moreover, these multiplier blocks under an architecture including a multiplier are locked by prominent logic locking methods, namely, RLL~\cite{roy08} and the \mbox{SAT-resilient} methods of AntiSAT~\cite{xie19}, SARLock~\cite{yasin16}, SFLL~\cite{yasin17}, CASLock~\cite{shakya19}, and SKGLock~\cite{nguyen21}. In this case, the multiplier block described at RTL is initially synthesized and its gate-level netlist is obtained, and then, this netlist is locked. Note that while the RLL, AntiSAT, and SFLL methods were applied using the NEOS tool~\cite{neos}, the script for the SARLock method was provided by P.~Subramanyan, and we implemented the CASLock and SKGLock methods. In the RLL method, 32 key bits are used, the same as the obfuscation techniques presented in Table~\ref{tab:obf_results}. Same as the hybrid protection method shown in Table~\ref{tab:obf_results}, there are a total of 48 key bits in the combination of RLL and a \mbox{SAT-resilient} method, while 16 key bits are designated to a \mbox{SAT-resilient} method. Note that due to the locking nature of AntiSAT, CASLock, and SKGLock, they require twice the number of designated key bits. Hence, a total of 32 key bits are used in these methods. Table~\ref{tab:lock_results} presents the results of locked multiplier blocks. The locked designs, whose secret key has not been discovered by the given attacks, are also highlighted.

\subsubsection{Comments on Hardware Complexity}

Observe from Table~\ref{tab:lock_results} that \mbox{SAT-resilient} methods with a combination of RLL lead to designs with hardware complexity very close to each other. When compared to the results of the hybrid protection technique given in Table~\ref{tab:obf_results} under the architectures using multiplier(s), the locked CAVM and MCM blocks have larger and smaller area, respectively and the locked TMCM blocks have competitive area. A locked MCM block has less hardware complexity than an obfuscated or protected MCM block because the logic locking is applied after the common subexpressions are exploited by the synthesis tool. 

\subsubsection{Comments on Attack Resiliency}

Also, observe from Table~\ref{tab:lock_results} that the secret key of designs locked by RLL, RLL+SARLock, and RLL+SFLL\footnote{Confirmed by the developer of the SFLL method that when a small number of key bits are used and their values are biased towards all logic 0s or 1s in the SFLL method, the exponential growth in the number of iterations in the \mbox{SAT-based} attack is no longer valid.} can be found by the given attacks. Note also that the MCM block locked by RLL+CASlock could be broken by the AppSAT. While the query attack is also capable of proving the values of most of the RLL key bits in all logic locking methods and extra SKGLock key bits, the SCOPE attack can predict the values of some key bits of designs locked by RLL+SARLock and RLL+SKGLock with high accuracy.

\subsection{Results of the Obfuscated and Locked FIR Filters}
\label{subsec:fir}

Table~\ref{tab:fir_results} presents the synthesis results of parallel direct and transposed form and folded FIR filters, whose CAVM, MCM, and TMCM blocks are obfuscated by the previously proposed techniques~\cite{aksoy21, pilato21} and the hybrid protection technique, respectively. It also shows the synthesis results of FIR filters locked by prominent logic locking methods. It introduces the results of attacks that can be applied to sequential circuits namely, the OG KC2 attack, which was taken from~\cite{neos}, and the OL SCOPE attack. In this table, \textit{failed} denotes that the found solution of the KC2 attack is actually a wrong key verified through simulation.

\subsubsection{Comments on Hardware Complexity}

Observe from Table~\ref{tab:fir_results} that the direct form filter has less area and consumes less power, but has a higher delay when compared to the transposed form filter. On the other hand, the folded design has the smallest area, but the filter output is computed in 30 clock cycles, increasing the latency and energy consumption. The conclusions drawn based on the gate-level synthesis results on the obfuscated and locked multipliers blocks given in Tables~\ref{tab:obf_results}-\ref{tab:lock_results} are also valid on the obfuscated and locked FIR filters. However, due to the registers in the FIR design, the overhead on the overall FIR filter design gets smaller. Note that the proposed hybrid technique achieves the maximum area reduction with respect to the logic locking methods on the parallel direct form FIR filters, i.e., 4.4\%, obtained when the FIR filter under the \mbox{\sc cavm-sa} architecture is compared to the FIR filter locked by RLL+SKGLock.

\subsubsection{Comments on Attack Resiliency}

Observe also from Table~\ref{tab:fir_results} that the KC2 attack is capable of discovering the secret key of the obfuscated FIR filters using previously proposed techniques, except for the direct form filters, but it is not successful on the FIR filters protected by the proposed hybrid technique. It can also find the secret key locked by RLL, except for the direct form filter, but fails on the filters locked by both RLL and a \mbox{SAT-resilient} method. Similarly, the SCOPE attack generally deciphers all the key bits on the obfuscated FIR filters with high accuracy, but it can only decipher a small number of key bits of the FIR filters protected by the hybrid technique. However, it is capable of deciphering more key bits on the locked FIR filters when compared to its results on the locked multiplier blocks. This is heavily due to the resynthesis of the FIR filter including the locked multiplier block. Note that the proposed hybrid technique increases the area and power dissipation of FIR filters when compared to the previously proposed obfuscation techniques~\cite{aksoy21, pilato21} in order to increase their resiliency to the existing attacks.

\begin{figure*}[t]
	\centering
	\vspace*{-4mm}
	\parbox{5.9cm}{\centerline{\includegraphics[width=6.5cm]{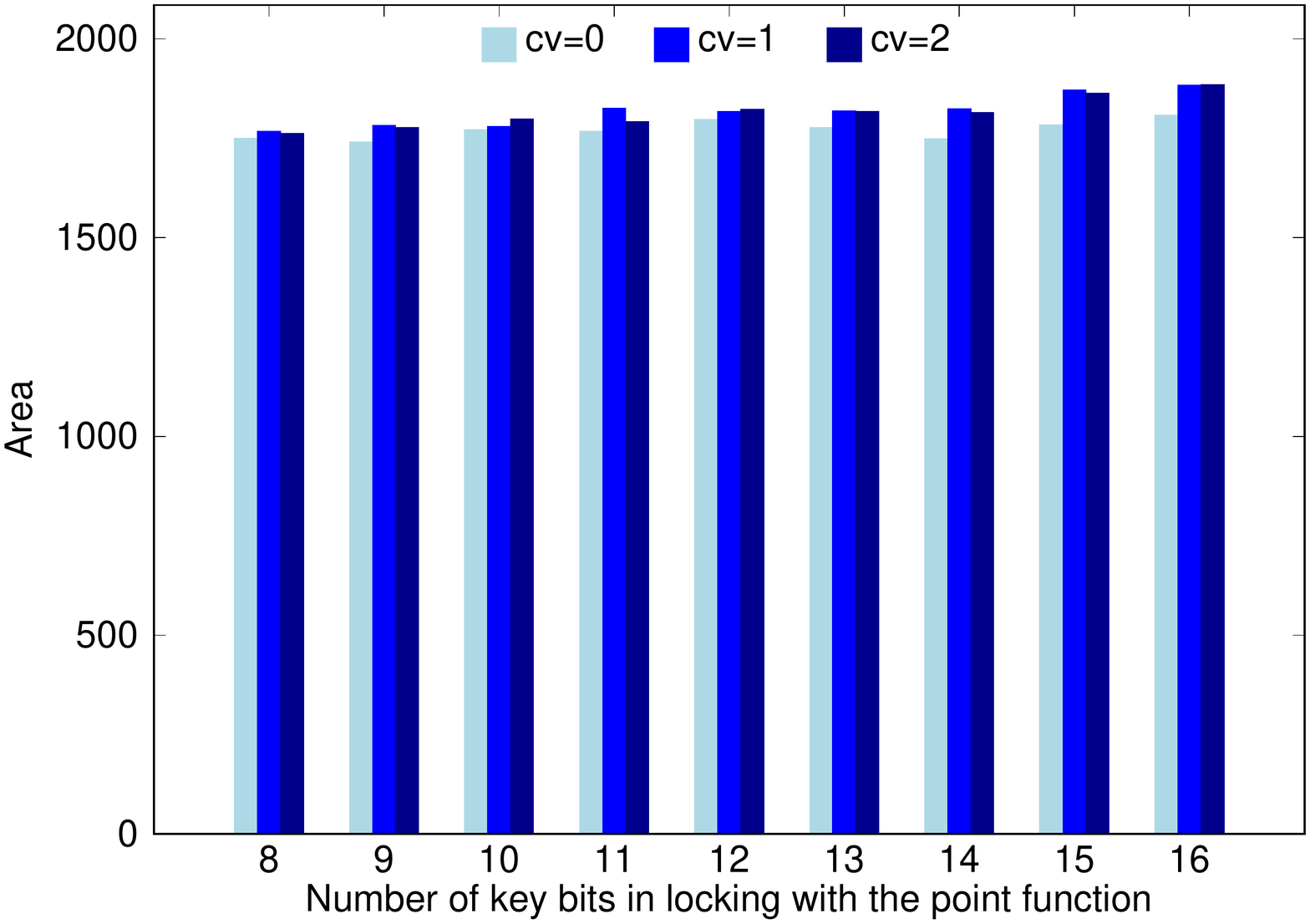}}}\
	\parbox{5.9cm}{\centerline{\includegraphics[width=6.5cm]{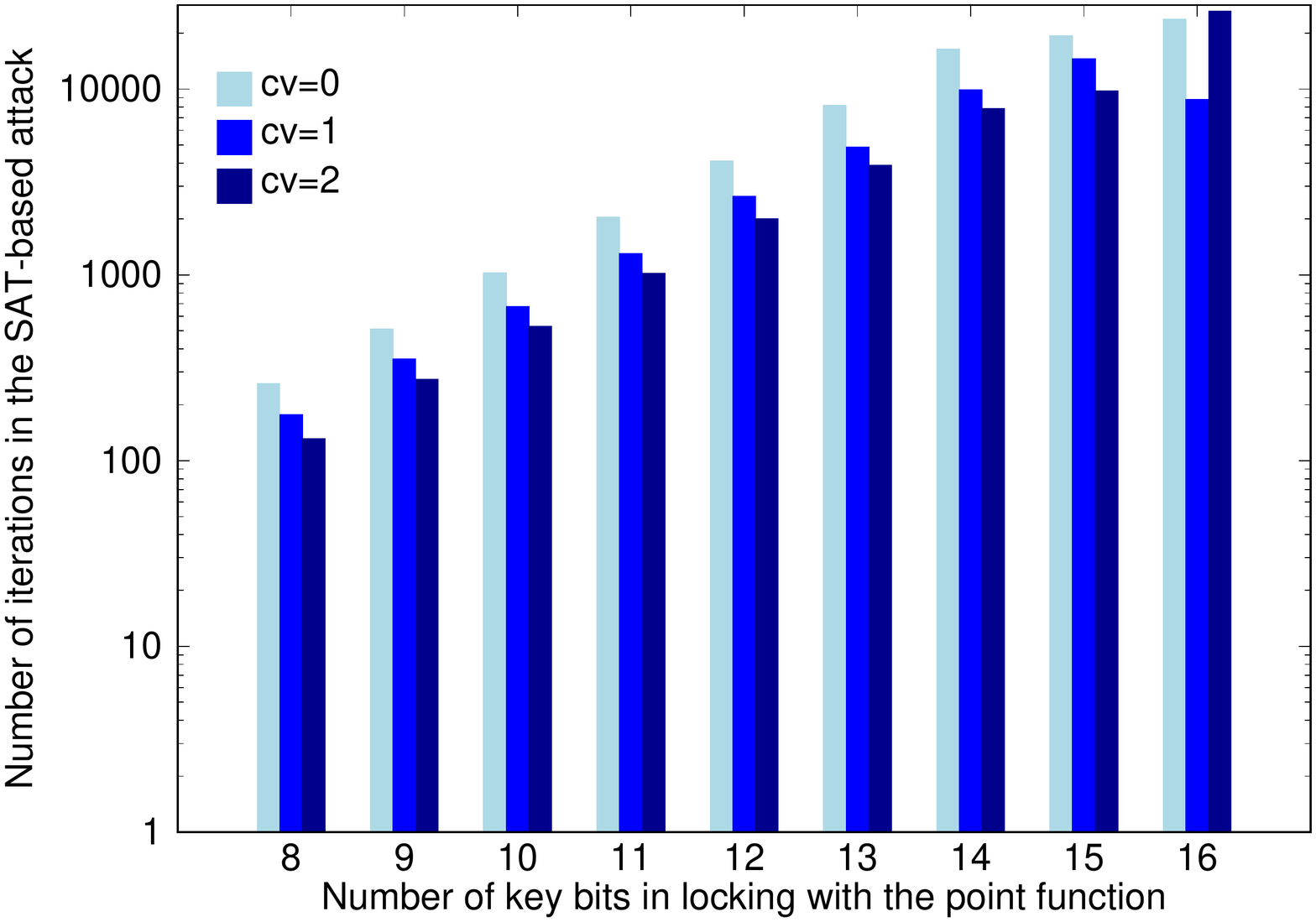}}}\
	\parbox{5.9cm}{\centerline{\includegraphics[width=6.5cm]{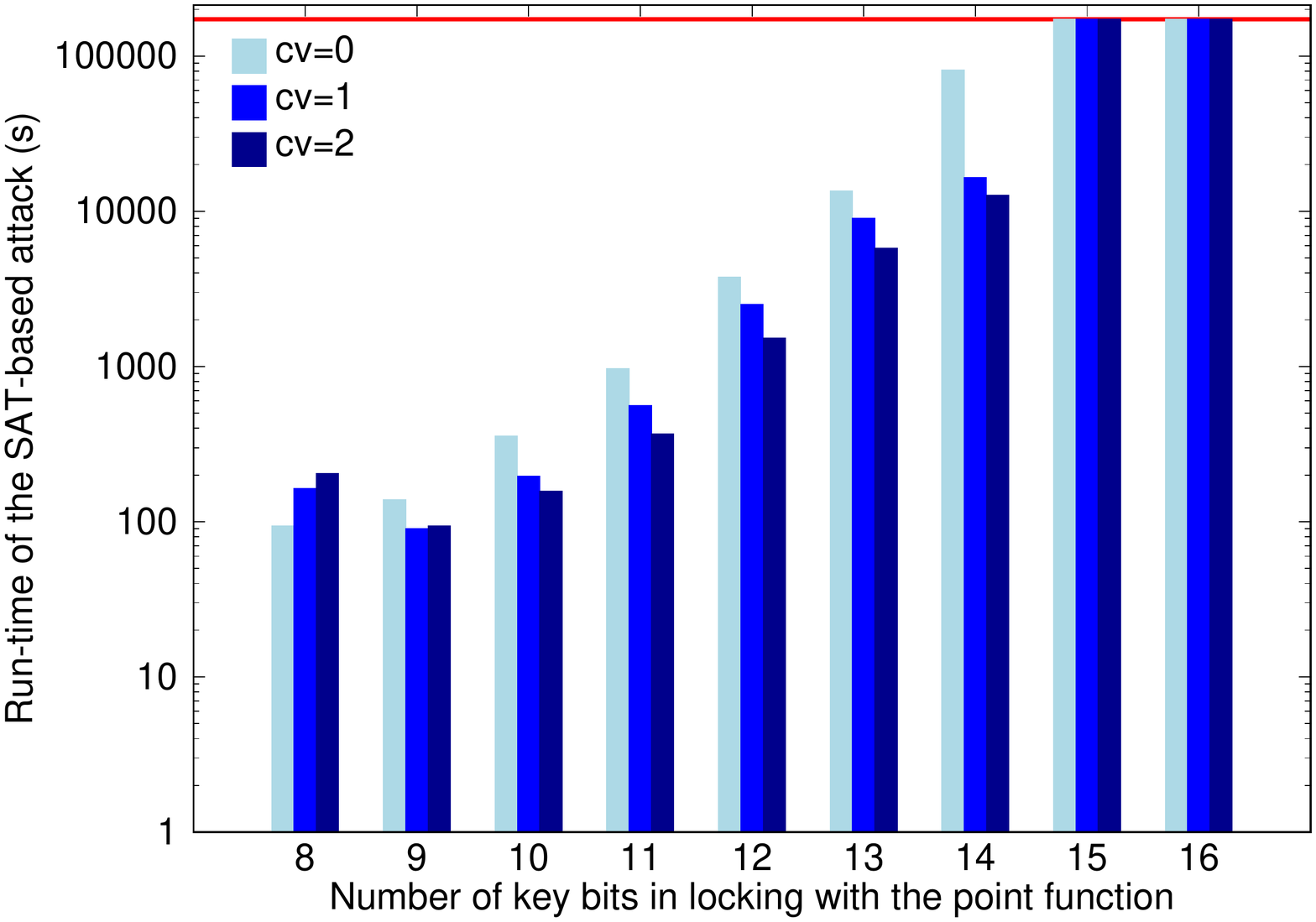}}}\
	
	\vspace*{-2mm}
	
	\parbox{5.9cm}{\centerline{\footnotesize (a)}}\
	\parbox{5.9cm}{\centerline{\footnotesize (b)}}\
	\parbox{5.9cm}{\centerline{\footnotesize (c)}}\
	\vspace*{-2mm}
	\caption{Impact of point function parameters: (a)~area; (b)~number of iterations; (c)~run-time.}  
	\label{fig:pf_results}
	\vspace*{-4mm}
\end{figure*}

\subsection{Analysis on the Point Function}
\label{subsec:analysis}

To find the impact of the point function and its parameters in the hybrid protection technique on the hardware complexity, the number of iterations taken in the SAT-based attack~\cite{subramanyan15}, and the run-time of the SAT-based attack~\cite{subramanyan15}, we used the TMCM block of our FIR filter under the {\sc tmcm-mul} architecture. Again, the TMCM block is obfuscated using decoys with 32 key bits when $ibw$ is 16. In logic locking with the point function, the number of key bits, i.e., $w$, is determined to be between 8 and 16, the corruption value, i.e., $cv$, is set to be between 0 and 2, and a single primary output is locked. Fig.~\ref{fig:pf_results} presents the impact of point function parameters on the area of the protected TMCM block and the number of iterations and run-time of the SAT-based attack.

Observe from Fig.~\ref{fig:pf_results}(a) that as the number of key bits used in logic locking, $w$, increases, the area of the protected TMCM block using the hybrid technique increases slightly. Note that as the corruption value, $cv$, increases, the design area increases simply due to the increased range of comparator logic given in Listing~\ref{lst:pf_code}. Also, observe from Figs.~\ref{fig:pf_results}(b)-(c) that as $w$ increases, the number of iterations and run-time of the SAT-based attack increases. An exponential growth in the number of iterations and run-time can be observed till $w$ is 15. As can be seen from Fig.~\ref{fig:pf_results}(c), for the 15- and 16-bit keys in the point function, the SAT-based attack cannot find the secret key in the time limit, i.e., 2 days, denoted by the red line. Thus, the number of iterations given in Fig.~\ref{fig:pf_results}(b) for these number of key bits, is the one obtained in the time limit. Note also that in all TMCM designs locked by the point function with the given parameters,  the number of iterations increases exponentially, while it decreases as $cv$ is increased, but still keeping the exponential growth. 

\begin{figure}[t]
	\centering
	\vspace*{-4mm}
	\parbox{8.5cm}{\vspace*{-2mm} \includegraphics[width=8.5cm]{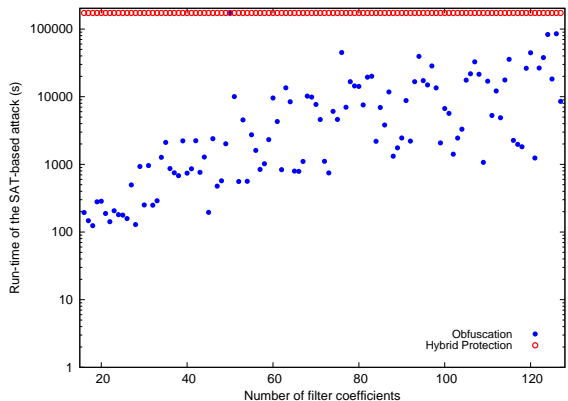}}
	\vspace*{-6mm}
	\caption{Run-time of the SAT-based attack on TMCM blocks.}
	\label{fig:sat_attack_random}
	\vspace*{-4mm}
\end{figure}

To find the impact of locking an obfuscated design using a point function on hardware complexity and attack resiliency, we used the same 112 FIR filters presented in Section~\ref{subsec:query_results}, where $n$ ranges between 16 and 127. In our experiments, the TMCM blocks of folded FIR filters were obfuscated using $2^{\lfloor log_2n \rfloor + 1}$ key bits under the {\sc tmcm-mul} architecture when $ibw$ was set to 16. For the point function, 16 key bits were used. Fig.~\ref{fig:sat_attack_random} presents the run-time of the SAT-based attack on obfuscated and protected TMCM blocks when its time limit was 2 days.

Observe from Fig.~\ref{fig:sat_attack_random} that locking an obfuscated design using a point function increases the SAT-based attack resiliency significantly. Note that the SAT-based attack can find a solution to all obfuscated TMCM blocks except one. However, the average area, delay, and power dissipation of the protected designs are increased by 10.7\%, 7.1\%, and 9.6\%, respectively when compared to those of the obfuscated designs.

\subsection{Analysis on the Direct Form FIR Filter}

Among the parallel design of FIR filters, the direct form is a good candidate to be used in a secure implementation for several reasons based on the results obtained in this work. First, as shown in Table~\ref{tab:fir_results}, its obfuscated hardware complexity in terms of area and power dissipation is significantly smaller than the transposed form filter. Second, it includes a large number of multiplication operations in chain, which make the SAT-based attack and its variants hard to discover the secret key. Third, the CAVM block of the direct form filter has a large number of inputs than the MCM block of the transposed form filter, which enables an increase in the number of key bits in the point function, improving the resiliency of the design protected by the hybrid technique as shown in Figs.~\ref{fig:pf_results}(b)-(c). This last observation is also true when compared to the TMCM block used in folded FIR filter design. 

\begin{table}[t]
	\centering
	\caption{Details on the protected and locked cavm blocks.} 
	\vspace*{-2mm}
	\begin{tabular}{|@{\hskip3pt}l@{\hskip3pt}|c@{\hskip3pt}c@{\hskip3pt}|c@{\hskip3pt}|c@{\hskip3pt}c@{\hskip3pt}c@{\hskip3pt}|c@{\hskip3pt}c@{\hskip3pt}c@{\hskip3pt}|}
		\hline
		\multicolumn{1}{|c|}{ \multirow{2}{*}{Filter} } & \multirow{2}{*}{$n$} & \multirow{2}{*}{$mbw$} & \multirow{2}{*}{$p$} & \multicolumn{3}{c|}{Hybrid} & \multicolumn{3}{c|}{RLL+CASLock} \\
		\cline{5-10}
		&                      &                        &                    & area & delay & power        & area & delay & power             \\
		\hline
		Dempster02   & 25  & 12 & 41  & 11145 & 5903 & 7057  & 16557 & 3136 & 4238 \\
		Johansson08  & 30  & 10 & 46  & 14826 & 4662 & 4504  & 16461 & 3381 & 4101 \\
		Shi11\_Y2    & 34  & 11 & 50  & 12270 & 3594 & 3567  & 12533 & 3163 & 3190 \\
		Shi11\_A     & 59  & 10 & 75  & 21431 & 4858 & 5825  & 20993 & 3404 & 5390 \\
		Samueli89    & 60  & 13 & 76  & 24188 & 4130 & 7522  & 26696 & 3998 & 6819 \\
		Lim83        & 63  & 10 & 79  & 22932 & 4848 & 6652  & 24696 & 3661 & 6322 \\
		Yoshino09    & 64  & 13 & 80  & 25096 & 4964 & 7927  & 30414 & 3437 & 7927 \\
		Nielsen89    & 67  & 15 & 83  & 26187 & 5318 & 9645  & 34950 & 4071 & 9164 \\
		Maskell07    & 108 & 9  & 124 & 37306 & 4811 & 9049  & 37148 & 3604 & 9522 \\
		LimYu07      & 121 & 14 & 137 & 48354 & 4947 & 16862 & 59492 & 3888 & 15346 \\
		\hline
	\end{tabular}
	\label{tab:dirfir}
	\vspace*{-6mm}
\end{table}

To find the impact of the number of coefficients on the hardware complexity of the protected CAVM block of an FIR filter, 10 filters were taken from~\cite{firsuite}, where $n$ ranges between 25 and 121 and $mbw$ is between 9 and 15. In the design of CAVM blocks, $ibw$ is set to 16. In the hybrid protection technique, these CAVM blocks were obfuscated using decoys with $n$ key bits and locked using the point function with $ibw$ key bits under the {\sc cavm-mul} architecture using a total of \mbox{$n+ibw$} key bits. These protected designs are also compared with those locked by both RLL and CASLock, where the number of RLL and CASLock key bits is $n-ibw$ and $2*ibw$, respectively. This logic locking method was chosen because it generally generates a locked multiplier block with a small area as shown in Table~\ref{tab:lock_results}. Table~\ref{tab:dirfir} presents the \mbox{gate-level} synthesis results of the CAVM designs protected by the hybrid technique and locked by both RLL and CASLock. 

Observe from Table~\ref{tab:dirfir} that as the number of coefficients, $n$, increases, the hardware complexity of the protected and locked CAVM blocks generally increases. The hybrid protection technique generally leads to a design with a smaller area when compared to the RLL+CASLock method, where the gain reaches up to 32.6\%. Note that on filters \textit{Shi11\_A} and \textit{Maskell07}, the RLL+CASLock method leads to a locked design with a smaller area, since the bit-width of coefficients is small, enabling the synthesis tool to optimize the logic.

\begin{figure}[t]
	\centering
	\vspace*{-4mm}
	\parbox{8.5cm}{\vspace*{-2mm} \includegraphics[width=8.5cm]{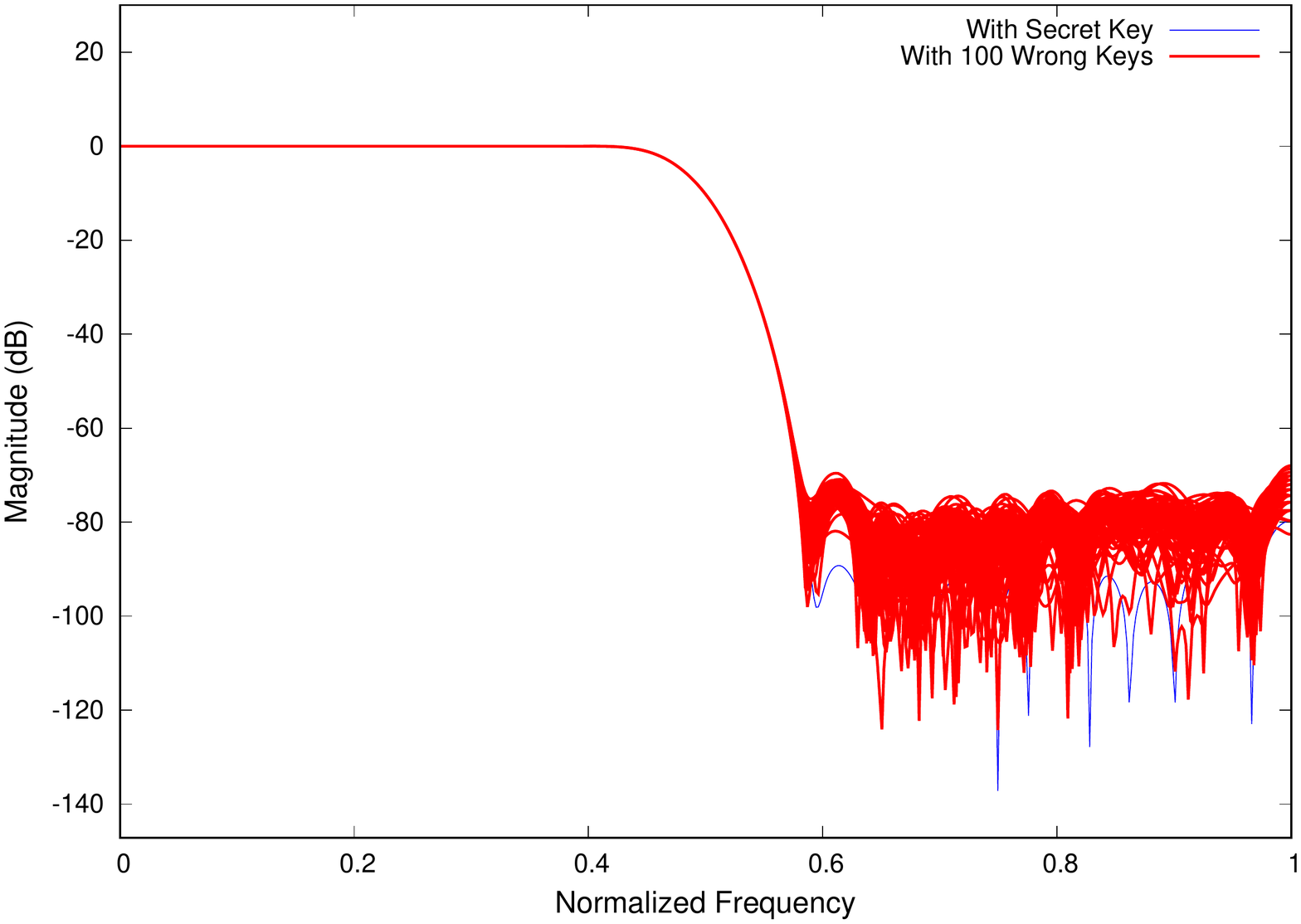}}
	\parbox{8.5cm}{\vspace*{-6mm} \centerline{\footnotesize (a)}}
	\parbox{8.5cm}{\vspace*{-8mm} \includegraphics[width=8.5cm]{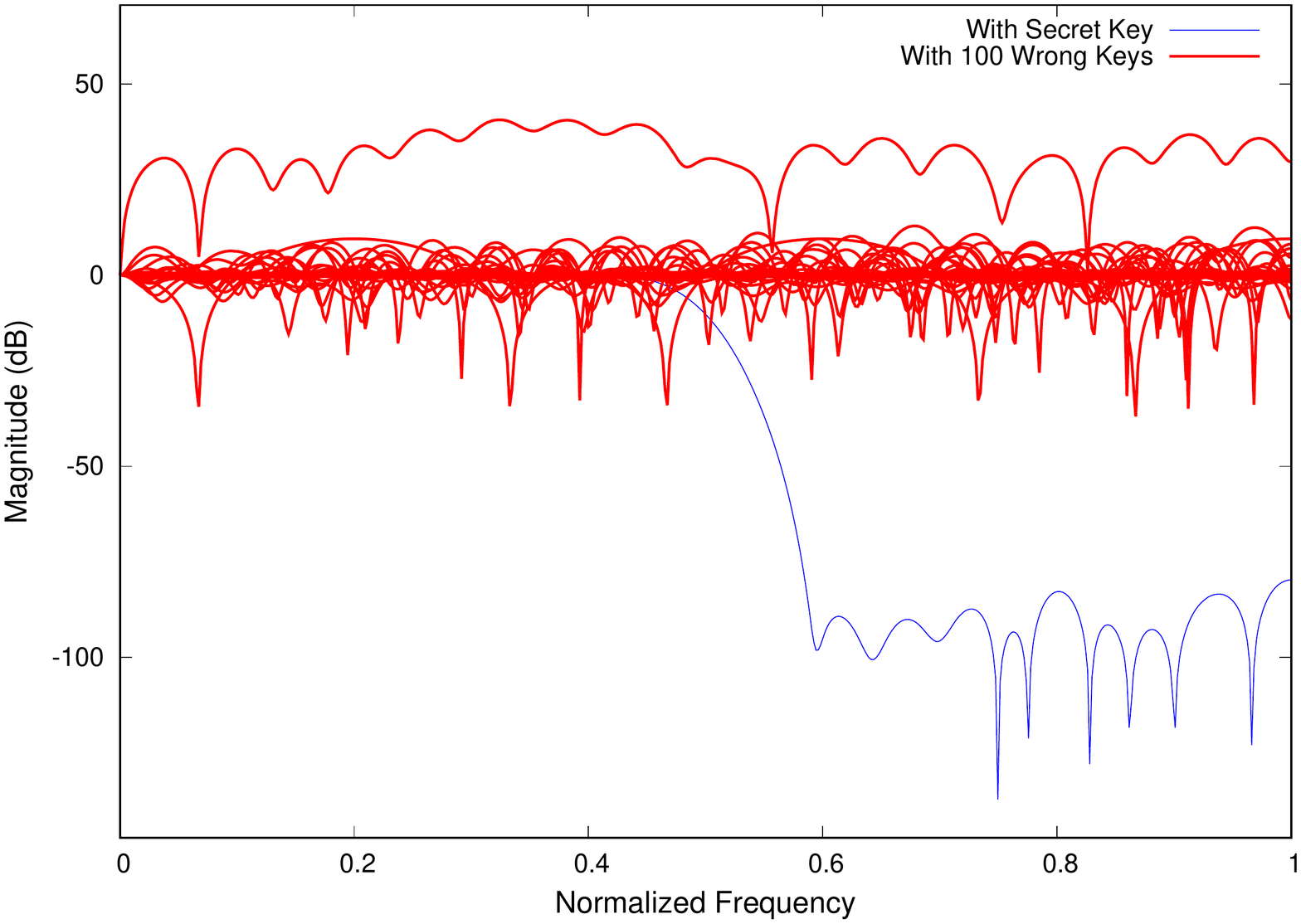}}
	\parbox{8.5cm}{\vspace*{-6mm} \centerline{\footnotesize (b)}}
	\vspace*{-6mm}
	\caption{Behavior of the FIR filter \textit{Nielsen89}: (a)~protected by the hybrid technique; (b)~locked by RLL+CASLock.}
	\label{fig:zpfr}
	\vspace*{-6mm}
\end{figure}

To find the impact of obfuscation techniques on the filter behavior, the Zero-Phase Frequency Response (ZPFR) of the FIR filter \textit{Nielsen89} is obtained when the secret key and 100 randomly generated wrong keys are applied. Fig.~\ref{fig:zpfr} presents ZPFRs of FIR filters protected by the hybrid technique and locked by RLL+CASlock. 

Observe from Fig.~\ref{fig:zpfr} that both obfuscation techniques may lead to a filter behavior different from the original one when a random wrong key is applied. While the filter behavior of the protected design under a wrong key is meaningful, but out of desired filter specification, the locked design exhibits an unmeaningful behavior under a wrong key due to the logic related to RLL key bits. Thus, the hybrid protection technique may make the adversary believe that the filter behavior under the wrong key is actually the desired one.

	\section{Discussion}
\label{sec:discussion}

Other than the logic locking attacks used in this article, there exist Reverse Engineering (RE) and Side-Channel Analysis (SCA) techniques that can identify the filter coefficients in an obfuscated design. In~\cite{aksoy22}, a machine learning tool that can determine the decoy selection method used in an obfuscated design was developed. The same work also proposed an RE technique that can identify filter coefficients hidden among decoys determined based on a decoy selection method. It was shown that if more than one decoy is used to obfuscate a filter coefficient, where the Hamming distance between each decoy and filter coefficient is 1, then the coefficient can be identified. In other cases, the RE technique was not capable of identifying original filter coefficients.

To the best of our knowledge, there exists no SCA technique proposed specifically to identify the original filter coefficients in an obfuscated filter design. The challenge for such a technique would be to understand how the synthesis tools embed the constants, i.e., filter coefficients and decoys, into the gate-level design using efficient methods, which optimize the hardware complexity of constant multiplications. This procedure would almost entail reverse engineering the algorithms used by the synthesis tools. In this case, it will be hard to reveal the filter coefficients from the power dissipation or delay values obtained from the obfuscated design since those data come from a logic combining both filter coefficients and decoys. Studying SCA and its efficiency to overcome obfuscation methods remains a formidable path for future research.

	\section{Conclusions}
\label{sec:conclusions}

This article focused on the obfuscation of digital FIR filters. Initially, it showed that the techniques previously proposed for the obfuscation of FIR filters are vulnerable to our \mbox{SAT-based} query attack, which applies several queries and proves that the found key bit value is the actual value of the related key bit in the secret key. Then, to secure an FIR filter design, it proposed the hybrid protection technique, which includes both obfuscation and locking with a point function. The proposed technique is applied to parallel direct and transposed forms of an FIR filter and its folded implementation. Experimental results clearly showed that the hybrid protection technique is competitive to prominent logic locking techniques in terms of hardware complexity and leads to obfuscated designs resilient to well-known attacks. It is also shown that the direct form FIR filter is a good candidate for secure filter implementation.

	\section*{Acknowledgment}

The authors would like to thank Nimisha Limaye and Satwik Patnaik for running our obfuscated designs on their tools and Mohammad Yasin, Leon Li, and Christian Pilato for fruitful discussions. The attacks were carried out in the High Performance Computing Centre of TalTech.

	\bibliographystyle{IEEEtran}
	\bibliography{tvlsi22}
	
	\begin{IEEEbiographynophoto}{Levent Aksoy}
	received his Ph.D. degree in electronics engineering from Istanbul Technical University (ITU), Istanbul, Türkiye, in 2009. He worked as a researcher at ITU and INESC-ID, Lisbon, Portugal. He also worked at Dialog Semiconductor, Istanbul, Türkiye, as a senior digital design engineer. Currently, he is a post-doctoral researcher at the Centre for Hardware Security at Tallinn University of Technology (TalTech), Tallinn, Estonia. His research interests include hardware security and CAD for VLSI circuits with emphasis on solving EDA problems using SAT models and optimization techniques.
\end{IEEEbiographynophoto}

\vspace*{-10mm}

\begin{IEEEbiographynophoto}{Quang-Linh Nguyen}
	currently works as a Design-for-Test engineer at STMicroelectronics, Grenoble, France. He received a M.S. degree in
	Integrated Circuits and Systems from University of Paris-Saclay, Paris, France, in 2018 and a PhD degree in Micro-Electronics from University of Montpellier, Montpellier, France, in 2022. His research interests include VLSI Design, Design-for-Trust, Design-for-Test and Hardware Security.
\end{IEEEbiographynophoto}

\vspace*{-10mm}

\begin{IEEEbiographynophoto}{Felipe Almeida} received his bachelor’s degree in Computer Engineering from the Pernambuco University and a master's degree in Microelectronics from the Federal University of Rio Grande do Sul. He is currently affiliated with the Centre for Hardware Security at Tallinn University of Technology (TalTech) as a Ph.D. student. His research interests are on Hardware Security and Radiation Tolerant Circuits.
\end{IEEEbiographynophoto}

\vspace*{-10mm}

\begin{IEEEbiographynophoto}{Jaan Raik}
	is a professor of digital systems' verification at the Department of Computer Systems and the head of the Centre for Dependable Computing Systems of TalTech University, Estonia. Prof. Raik received his M.Sc. and Ph.D. degrees at TalTech in 1997 and in 2001, respectively. He has co-authored more than 200 peer-reviewed scientific publications. His research interests cover a wide area in electrical engineering and computer science domains including hardware test, functional verification, fault-tolerance and security as well as emerging computer architectures. 
\end{IEEEbiographynophoto}

\vspace*{-10mm}

\begin{IEEEbiographynophoto}{Marie-Lise Flottes}
	is a researcher for the French National Research Center (CNRS). Since 1990, she has been conducting research at LIRMM, Montpellier, France. She received her Ph.D. degree in 1990 from the University of Montpellier. Her interests include design for testability, testability and dependability of secure circuits, test data compression and test management for SoC and SiP.
\end{IEEEbiographynophoto}

\vspace*{-10mm}

\begin{IEEEbiographynophoto}{Sophie Dupuis}
	has been an Associate Professor with LIRMM, Montpellier, France, since 2011. She received her M.Sc. and Ph.D. degrees in micro-electronics and design on integrated circuits from the Pierre \& Marie Curie University, Paris, France, in 2004 and 2009 respectively. Her current research interests are oriented towards hardware trust, the design of trusted circuits despite potential untrustworthy design steps in particular.
\end{IEEEbiographynophoto}

\vspace*{-10mm}

\begin{IEEEbiographynophoto}{Samuel Pagliarini} received the PhD degree from Telecom ParisTech, Paris, France, in 2013. He has held research positions with the University of Bristol, Bristol, UK, and with Carnegie Mellon University, Pittsburgh, PA, USA. He is currently a Professor at Tallinn University of Technology (TalTech) in Tallinn, Estonia where he leads the Centre for Hardware Security. His current research interests include many facets of digital circuit design, with a focus on circuit reliability, dependability, and hardware trustworthiness.
\end{IEEEbiographynophoto}

\end{document}